\documentclass[aps,prd,twocolumn,showpacs,preprintnumbers,amsmath,amssymb,superscriptaddress,altaffilletter,reprint]{revtex4-1}

\usepackage{graphicx}
\usepackage{hyperref}

\begin{document}

\title{Measurement of inclusive charged current interactions on carbon in a few-GeV neutrino beam}


\affiliation{Institut de Fisica d'Altes Energies, Universitat Autonoma de Barcelona, E-08193 Bellaterra (Barcelona), Spain}
\affiliation{Department of Physics, University of Colorado, Boulder, Colorado 80309, USA}
\affiliation{Department of Physics, Columbia University, New York, NY 10027, USA}
\affiliation{Fermi National Accelerator Laboratory; Batavia, IL 60510, USA}
\affiliation{High Energy Accelerator Research Organization (KEK), Tsukuba, Ibaraki 305-0801, Japan}
\affiliation{Department of Physics, Imperial College London, London SW7 2AZ, UK}
\affiliation{Department of Physics, Indiana University, Bloomington, IN 47405, USA}
\affiliation{Kamioka Observatory, Institute for Cosmic Ray Research, University of Tokyo, Gifu 506-1205, Japan}
\affiliation{Research Center for Cosmic Neutrinos, Institute for Cosmic Ray Research, University of Tokyo, Kashiwa, Chiba 277-8582, Japan}
\affiliation{Department of Physics, Kyoto University, Kyoto 606-8502, Japan}
\affiliation{Los Alamos National Laboratory; Los Alamos, NM 87545, USA}
\affiliation{Department of Physics and Astronomy, Louisiana State University, Baton Rouge, LA 70803, USA}
\affiliation{Department of Physics, Massachusetts Institute of Technology, Cambridge, MA 02139, USA}
\affiliation{Department of Chemistry and Physics, Purdue University Calumet, Hammond, IN 46323, USA}
\affiliation{Universit$\grave{a}$ di Roma La Sapienza, Dipartimento di Fisica and INFN, I-00185 Rome, Italy}
\affiliation{Physics Department, Saint Mary's University of Minnesota, Winona, MN 55987, USA}
\affiliation{Department of Physics, Tokyo Institute of Technology, Tokyo 152-8551, Japan}
\affiliation{Instituto de Fisica Corpuscular, Universidad de Valencia and CSIC, E-46071 Valencia, Spain}

\author{Y.~Nakajima}\affiliation{Department of Physics, Kyoto University, Kyoto 606-8502, Japan}
\author{J.~L.~Alcaraz-Aunion}\affiliation{Institut de Fisica d'Altes Energies, Universitat Autonoma de Barcelona, E-08193 Bellaterra (Barcelona), Spain}
\author{S.~J.~Brice}\affiliation{Fermi National Accelerator Laboratory; Batavia, IL 60510, USA}
\author{L.~Bugel}\affiliation{Department of Physics, Massachusetts Institute of Technology, Cambridge, MA 02139, USA}
\author{J.~Catala-Perez}\affiliation{Instituto de Fisica Corpuscular, Universidad de Valencia and CSIC, E-46071 Valencia, Spain}
\author{G.~Cheng}\affiliation{Department of Physics, Columbia University, New York, NY 10027, USA}
\author{J.~M.~Conrad}\affiliation{Department of Physics, Massachusetts Institute of Technology, Cambridge, MA 02139, USA}
\author{Z.~Djurcic}\affiliation{Department of Physics, Columbia University, New York, NY 10027, USA}
\author{U.~Dore}\affiliation{Universit$\grave{a}$ di Roma La Sapienza, Dipartimento di Fisica and INFN, I-00185 Rome, Italy}
\author{D.~A.~Finley}\affiliation{Fermi National Accelerator Laboratory; Batavia, IL 60510, USA}
\author{A.~J.~Franke}\affiliation{Department of Physics, Columbia University, New York, NY 10027, USA}
\author{C.~Giganti}\altaffiliation[Present address: ]{DSM/Irfu/SPP, CEA Saclay, F-91191 Gif-sur-Yvette, France}\affiliation{Universit$\grave{a}$ di Roma La Sapienza, Dipartimento di Fisica and INFN, I-00185 Rome, Italy}
\author{J.~J.~Gomez-Cadenas}\affiliation{Instituto de Fisica Corpuscular, Universidad de Valencia and CSIC, E-46071 Valencia, Spain}
\author{P.~Guzowski}\affiliation{Department of Physics, Imperial College London, London SW7 2AZ, UK}
\author{A.~Hanson}\affiliation{Department of Physics, Indiana University, Bloomington, IN 47405, USA}
\author{Y.~Hayato}\affiliation{Kamioka Observatory, Institute for Cosmic Ray Research, University of Tokyo, Gifu 506-1205, Japan}
\author{K.~Hiraide}\affiliation{Department of Physics, Kyoto University, Kyoto 606-8502, Japan}\affiliation{Kamioka Observatory, Institute for Cosmic Ray Research, University of Tokyo, Gifu 506-1205, Japan}
\author{G.~Jover-Manas}\affiliation{Institut de Fisica d'Altes Energies, Universitat Autonoma de Barcelona, E-08193 Bellaterra (Barcelona), Spain}
\author{G.~Karagiorgi}\affiliation{Department of Physics, Massachusetts Institute of Technology, Cambridge, MA 02139, USA}
\author{T.~Katori}\affiliation{Department of Physics, Indiana University, Bloomington, IN 47405, USA}
\affiliation{Department of Physics, Massachusetts Institute of Technology, Cambridge, MA 02139, USA}
\author{Y.~K.~Kobayashi}\affiliation{Department of Physics, Tokyo Institute of Technology, Tokyo 152-8551, Japan}
\author{T.~Kobilarcik}\affiliation{Fermi National Accelerator Laboratory; Batavia, IL 60510, USA}
\author{H.~Kubo}\affiliation{Department of Physics, Kyoto University, Kyoto 606-8502, Japan}
\author{Y.~Kurimoto}\affiliation{Department of Physics, Kyoto University, Kyoto 606-8502, Japan}\affiliation{High Energy Accelerator Research Organization (KEK), Tsukuba, Ibaraki 305-0801, Japan}
\author{W.~C.~Louis}\affiliation{Los Alamos National Laboratory; Los Alamos, NM 87545, USA}
\author{P.~F.~Loverre}\affiliation{Universit$\grave{a}$ di Roma La Sapienza, Dipartimento di Fisica and INFN, I-00185 Rome, Italy}
\author{L.~Ludovici}\affiliation{Universit$\grave{a}$ di Roma La Sapienza, Dipartimento di Fisica and INFN, I-00185 Rome, Italy}
\author{K.~B.~M.~Mahn}\altaffiliation[Present address: ]{TRIUMF, Vancouver, British Columbia, V6T 2A3, Canada}\affiliation{Department of Physics, Columbia University, New York, NY 10027, USA}
\author{C.~Mariani}\affiliation{Department of Physics, Columbia University, New York, NY 10027, USA}
\author{S.~Masuike}\affiliation{Department of Physics, Tokyo Institute of Technology, Tokyo 152-8551, Japan}
\author{K.~Matsuoka}\affiliation{Department of Physics, Kyoto University, Kyoto 606-8502, Japan}
\author{V.~T.~McGary}\affiliation{Department of Physics, Massachusetts Institute of Technology, Cambridge, MA 02139, USA}
\author{W.~Metcalf}\affiliation{Department of Physics and Astronomy, Louisiana State University, Baton Rouge, LA 70803, USA}
\author{G.~B.~Mills}\affiliation{Los Alamos National Laboratory; Los Alamos, NM 87545, USA}
\author{G.~Mitsuka}\altaffiliation[Present address: ]{Solar-Terrestrial Environment Laboratory, Nagoya University, Furo-cho, Chikusa-ku, Nagoya, Japan}\affiliation{Research Center for Cosmic Neutrinos, Institute for Cosmic Ray Research, University of Tokyo, Kashiwa, Chiba 277-8582, Japan}
\author{Y.~Miyachi}\altaffiliation[Present address: ] {Department of Physics, Yamagata University, Yamagata, 990-8560 Japan}\affiliation{Department of Physics, Tokyo Institute of Technology, Tokyo 152-8551, Japan}
\author{S.~Mizugashira}\affiliation{Department of Physics, Tokyo Institute of Technology, Tokyo 152-8551, Japan}
\author{C.~D.~Moore}\affiliation{Fermi National Accelerator Laboratory; Batavia, IL 60510, USA}
\author{T.~Nakaya}\affiliation{Department of Physics, Kyoto University, Kyoto 606-8502, Japan}
\author{R.~Napora}\affiliation{Department of Chemistry and Physics, Purdue University Calumet, Hammond, IN 46323, USA}
\author{P.~Nienaber}\affiliation{Physics Department, Saint Mary's University of Minnesota, Winona, MN 55987, USA}
\author{D.~Orme}\affiliation{Department of Physics, Kyoto University, Kyoto 606-8502, Japan}
\author{M.~Otani}\affiliation{Department of Physics, Kyoto University, Kyoto 606-8502, Japan}
\author{A.~D.~Russell}\affiliation{Fermi National Accelerator Laboratory; Batavia, IL 60510, USA}
\author{F.~Sanchez}\affiliation{Institut de Fisica d'Altes Energies, Universitat Autonoma de Barcelona, E-08193 Bellaterra (Barcelona), Spain}
 \author{M.~H.~Shaevitz}\affiliation{Department of Physics, Columbia University, New York, NY 10027, USA}
\author{T.-A.~Shibata}\affiliation{Department of Physics, Tokyo Institute of Technology, Tokyo 152-8551, Japan}
\author{M.~Sorel}\affiliation{Instituto de Fisica Corpuscular, Universidad de Valencia and CSIC, E-46071 Valencia, Spain}
\author{R.~J.~Stefanski}\affiliation{Fermi National Accelerator Laboratory; Batavia, IL 60510, USA}
\author{H.~Takei}\altaffiliation[Present address: ]{Kitasato University, Tokyo, 108-8641 Japan}\affiliation{Department of Physics, Tokyo Institute of Technology, Tokyo 152-8551, Japan}
\author{H.-K.~Tanaka}\altaffiliation[Present address: ]{Brookhaven National Laboratory, Upton, NY 11973, USA}\affiliation{Department of Physics, Massachusetts Institute of Technology, Cambridge, MA 02139, USA}
\author{M.~Tanaka}\affiliation{High Energy Accelerator Research Organization (KEK), Tsukuba, Ibaraki 305-0801, Japan}
\author{R.~Tayloe}\affiliation{Department of Physics, Indiana University, Bloomington, IN 47405, USA}
\author{I.~J.~Taylor}\altaffiliation[Present address: ]{Department of Physics and Astronomy, State University of New York, Stony Brook, NY 11794-3800, USA }\affiliation{Department of Physics, Imperial College London, London SW7 2AZ, UK}
\author{R.~J.~Tesarek}\affiliation{Fermi National Accelerator Laboratory; Batavia, IL 60510, USA}
\author{Y.~Uchida}\affiliation{Department of Physics, Imperial College London, London SW7 2AZ, UK}
\author{R.~Van~de~Water}\affiliation{Los Alamos National Laboratory; Los Alamos, NM 87545, USA}
\author{J.~J.~Walding}\altaffiliation[Present address: ]{Department of Physics, College of William \& Mary, Williamsburg, VA 23187, USA}\affiliation{Department of Physics, Imperial College London, London SW7 2AZ, UK}
\author{M.~O.~Wascko}\affiliation{Department of Physics, Imperial College London, London SW7 2AZ, UK}
\author{H.~B.~White}\affiliation{Fermi National Accelerator Laboratory; Batavia, IL 60510, USA}
\author{M.~Yokoyama}\altaffiliation[Present address: ]{Department of Physics, University of Tokyo, Tokyo 113-0033, Japan}\affiliation{Department of Physics, Kyoto University, Kyoto 606-8502, Japan}
\author{G.~P.~Zeller}\affiliation{Fermi National Accelerator Laboratory; Batavia, IL 60510, USA}
\author{E.~D.~Zimmerman}\affiliation{Department of Physics, University of Colorado, Boulder, Colorado 80309, USA}
\collaboration{SciBooNE Collaboration}\noaffiliation


\begin{abstract}
The SciBooNE Collaboration reports a measurement of inclusive charged current interactions
of muon neutrinos on carbon with an average energy of 0.8~GeV using the Fermilab Booster Neutrino Beam.
We compare our measurement with two neutrino interaction simulations: NEUT and NUANCE.
The charged current interaction rates (product of flux and cross section) 
are extracted 
by fitting the muon kinematics,
with a precision of 6-15\% for the energy dependent 
and 3\% for the energy integrated analyses.
We also extract 
CC inclusive interaction cross sections from the observed rates, 
with a precision of  10-30\% for the energy dependent 
and 8\% for the energy integrated analyses.
This is the first measurement of the CC inclusive cross section on carbon around 1 GeV.
These results can be used to convert previous SciBooNE cross section ratio measurements 
to absolute cross section values.

\end{abstract}


\date{\today}
\pacs{13.15.+g, 25.30.Pt}
\preprint{FERMILAB-PUB-10-448-E}

\maketitle


\section{Introduction}
\label{sec:introduction}

The neutrino charged current (CC) interaction is the process most
commonly used to measure neutrino-nucleus scattering in the few-GeV 
region.  It is important for neutrino oscillation measurements because
the neutral current (NC) interaction is flavor-blind.  However, this
interaction mode is poorly understood because of large neutrino flux
and cross section uncertainties.

In the energy region above $\sim$3 GeV, 
NOMAD~\cite{:2007rv} and MINOS~\cite{Adamson:2009ju} recently reported
precise CC inclusive interaction cross section measurements.
However, in the $\sim$1~GeV region, all inclusive CC 
measurements have been made on deuterium targets using 
bubble chambers \cite{Barish:1978pj,Baker:1982ty}.
In this energy region, the nuclear effects of the neutrino target 
material (from Fermi motion and the nuclear potential) are significant.
Therefore, the cross section on deuterium targets is not 
directly applicable to the heavier nuclear target materials
used in the recent accelerator-based neutrino experiments in this energy region, 
e.g. SciBooNE, MiniBooNE~\cite{AguilarArevalo:2007it} and T2K~\cite{Itow:2001ee}.
Furthermore, the results in Ref.~\cite{Baker:1982ty} use CC quasi elastic (QE)
interactions to normalize the absolute neutrino flux,
which introduces additional ambiguity from the choice of 
CC-QE interaction model parameters.

Because of the poor knowledge of the cross section, 
there are multiple neutrino interaction simulators
used to predict different cross sections and kinematics of 
final state particles.
Among these, NEUT~\cite{Hayato:2002sd,Mitsuka:2008zz} 
and NUANCE~\cite{Casper:2002sd} 
are commonly used for the recent neutrino oscillation and interaction measurements.
NEUT is used  in the Kamiokande~\cite{Fukuda:1994mc}, 
Super-Kamiokande~\cite{Ashie:2005ik}, 
K2K~\cite{Ahn:2006zza}, and T2K experiments, 
while NUANCE is used in MiniBooNE and as a check of simulations by 
multiple experiments.
They are both tuned to describe the experimental data, 
however, they have not yet been precisely compared with each other in a single experiment.

In this paper, we report on the comparison of CC candidate events 
observed in SciBooNE with predictions based on the NEUT and NUANCE 
neutrino interaction simulators, as well as on the measured CC 
interaction rate (product of flux and cross section) as 
a function of neutrino energy extracted from these comparisons. 
We also report on 
the first measurement of the CC inclusive interaction cross section
on carbon in the 1 GeV region, which is relevant to 
ongoing and future neutrino oscillation experiments.

\begin{figure}[bhtp]
  \includegraphics[width=\columnwidth]{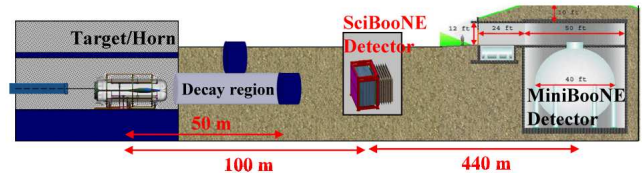}
  \caption{(color online). 
    Schematic overview of the Booster Neutrino Beamline 
    and the location of the SciBooNE and the MiniBooNE detectors. 
    (not to scale).
  }
  \label{fig:sciboone_setup}
\end{figure}

In addition to  this general purpose,
this analysis is also motivated by 
two direct applications to our  measurements.
The first 
is to provide a constraint on the product of  flux and  cross-section for
the forthcoming SciBooNE and MiniBooNE joint $\nu_\mu$ disappearance analysis~\cite{Nakajima:2010wc}.
The MiniBooNE collaboration recently performed a search for $\nu_\mu$ and 
$\overline{\nu}_\mu$ disappearance using only MiniBooNE data,
in which the largest uncertainty stems from the 
flux and cross section uncertainties~\cite{AguilarArevalo:2009yj}.
As shown in Fig.~\ref{fig:sciboone_setup}, the SciBooNE detector 
is located 440~m upstream of the MiniBooNE detector, sharing the 
same neutrino beam.
In addition, the neutrino target materials are both 
essentially carbon;  polystyrene (C${}_8$H${}_8$) for SciBooNE,
and mineral oil (CH${}_2$) for MiniBooNE.
Therefore, 
most of the flux and cross section uncertainties cancel when we compare the data
from the two experiments.
Then, a $\nu_\mu$ disappearance search with 
higher sensitivity becomes possible~\cite{Nakajima:2010wc}.

The second motivation
is to provide an absolute normalization factor for SciBooNE's previous 
cross section ratio measurements.
The SciBooNE collaboration recently measured relative cross sections of 
charged current coherent pion production~\cite{Hiraide:2008eu} and neutral current
neutral pion production~\cite{Kurimoto:2009wq,Kurimoto:2010rc},
which were both 
normalized to CC inclusive interactions.
This paper provides an absolute CC interaction cross section 
with the same definition used in the previous analyses, 
so that the measured ratios can be converted to absolute cross sections.

The remainder of this paper is organized as follows:
Section~\ref{sec:experiment} summarizes
the SciBooNE experiment, including the Booster
Neutrino Beamline (BNB) and the SciBooNE detector.
The simulation of neutrino interactions with nuclei is described in 
Section~\ref{sec:neutrinointeractionmc}.
Section~\ref{sec:reconstruction} describes the methods used to reconstruct 
CC interactions and defines the sub-samples used.
The analysis of the energy dependent neutrino interaction rate
is described in Section~\ref{sec:spectrum_fit}.
Finally, the results of the CC interaction rate analysis and the extracted 
CC absolute cross section are presented in Section~\ref{sec:results}.
The final conclusions are given in Section~\ref{sec:conclusions}.


\section{SciBooNE Experiment}
\label{sec:experiment}

\subsection{Neutrino Beam}
\label{sec:beam}

SciBooNE detected neutrinos produced by the Fermilab BNB. The same BNB
beam is also serving the MiniBooNE experiment. The BNB uses protons
accelerated to 8~GeV kinetic energy by the Fermilab Booster
synchrotron.  Beam properties are monitored on a spill-by-spill basis,
and at various locations along the BNB line. Transverse and
directional alignment of the beam, beam width and angular divergence,
beam intensity and losses along the BNB, are measured and used in the
data quality selection\cite{Hiraide:2008eu}. Protons strike a 71.1~cm 
long beryllium
target, producing a secondary beam of hadrons, mainly pions with a
small fraction of kaons. A cylindrical horn electromagnet made of
aluminum surrounds the beryllium target to sign-select and focus the
secondary beam.  For the data set used in this measurement, the horn
polarity was set to neutrino mode, focusing particles with positive
electric charge. The neutrino beam is mostly produced in the 50~m long
decay region, mainly from $\pi^+\to\mu^+ \nu_{\mu}$ in-flight decays.

The analysis presented here uses the full neutrino data set:
0.99$\times$10$^{20}$ protons on target (POT) collected between
October 2007 and April 2008.

\subsection{Neutrino Flux Prediction}
\label{sec:flux}

Predictions for the BNB neutrino flux illuminating the SciBooNE
detector are obtained via a GEANT4~\cite{Agostinelli:2002hh}
simulation of the beamline.  Hadronic interactions in the beryllium
target are carefully modeled.  For $\pi^+$ production, a
parametrization based on HARP~\cite{:2007gt} and BNL
E910~\cite{:2007nb} data are used. 
The total, inelastic and  quasi-elastic hadronic cross sections 
of protons, neutrons, $\pi^+$s and $\pi^-$s with Be or Al are treated via custom
cross section models~\cite{AguilarArevalo:2008yp}.
Other hadronic and all
electromagnetic processes of importance to neutrino production are
described by standard GEANT4 models.  For a detailed
description of the BNB simulation code, see Ref.
\cite{AguilarArevalo:2008yp}.  A total neutrino flux per proton on
target of $2.2\times 10^{-8}$~cm$^{-2}$ is expected at the SciBooNE
detector location for the neutrino running mode.  The flux is
dominated by muon neutrinos (93\%).

\begin{figure}[bhtp]
  \includegraphics[width=\columnwidth]{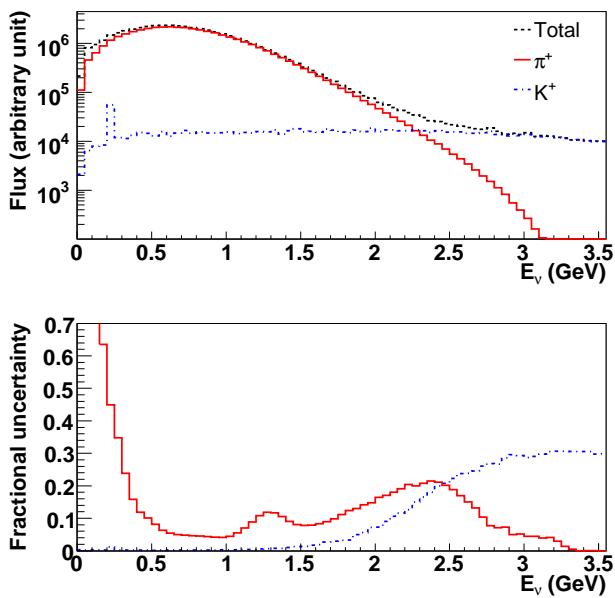}
  \caption{(color online).
    (Top) $\nu_\mu$ flux prediction at the SciBooNE detector as a
    function of neutrino energy $E_{\nu}$.
    The total flux and contributions from $\pi^+$ and $K^+$ decays 
    are shown.
    (Bottom) Fractional uncertainty of the $\nu_\mu$ flux prediction
    due to $\pi^+$ and $K^+$ production from the p-Be interaction.
     Additional uncertainties from POT, hadron interactions in the target,
    and the horn magnetic field are not shown.
  }
  \label{fig:flux-parents}
\end{figure}

Figure~\ref{fig:flux-parents} shows the $\nu_\mu$ flux from
$\pi^+$ and $K^+$ decays, and their fractional uncertainties.
The $\nu_\mu$ energy spectrum peaks at $\sim$ 0.6~GeV, and extends up to 2-3~GeV.
The mean  energy of the $\nu_\mu$ flux
is expected to be 0.76~GeV.
The largest error on the predicted neutrino flux results 
from the uncertainty of pion production in the initial p-Be 
process in the target. The simulation predicts that 96.7\% 
of muon neutrinos in the BNB are produced via $\pi^+$ decay. 
The uncertainty in 
$\pi^+$ production is determined from spline fits to the HARP 
$\pi^+$ double differential cross section data \cite{AguilarArevalo:2008yp}. 
The HARP data used are those from a thin (5\% 
interaction length) beryllium target run \cite{:2007gt}. While the HARP
data provide a valuable constraint on the BNB flux prediction, 
additional uncertainties resulting from thick target 
effects (secondary re-scattering of protons and pions) are 
included through the BNB flux simulation. 
The resulting $\pi^+$ production uncertainty is $\approx$ 5\% at 
the peak of the flux distribution and increases significantly at high
and low neutrino energies. 

The flux from $K^+$ decay is dominant for $E_\nu > 2.3$~GeV.  Since no
published data exist for $K^+$ production at the BNB primary proton
beam energy, we employ the Feynman scaling hypothesis to relate
$K^+$ production measurements at different proton beam energies to the
expected production at the BNB proton beam energy
\cite{AguilarArevalo:2008yp}.  The errors of the Feynman scaling
parameters obtained from these measurements are accounted for in the
systematic error.  Other major contributions to the flux error include
uncertainties on hadron interactions in the target and simulation
of the the horn magnetic field, which both contribute to shape and
normalization uncertainties, as well as the measurement of the number
of POT, which is a pure normalization uncertainty.  All flux errors
are modeled through variations in the simulation and result in a total
error of $\approx$ 7\% at the peak of the flux.

\subsection{SciBooNE Detector}
\label{sec:detector}

The SciBooNE detector is located 100~m downstream from the beryllium
target on the beam 
axis, as shown in Fig.~\ref{fig:sciboone_setup}.
The detector comprises three
sub-detectors: a fully active and finely segmented scintillator
tracker (SciBar), an electromagnetic calorimeter (EC), and a muon
range detector (MRD).  SciBar is the primary neutrino target
for this analysis.
In this analysis, we use the data from SciBar and the MRD, 
and the data from the EC are not used.

SciBooNE uses a right-handed Cartesian coordinate system in which the
$z$ axis is the beam direction and the $y$ axis is the vertical upward
direction.  The origin is located on the most upstream surface of
SciBar in the $z$ dimension, and at the center of the SciBar
scintillator plane in the $x$ and $y$ dimensions.  Since each
sub-detector is read out both vertically and horizontally, two views
are defined: top ($x$ vs. $z$ projection) and side ($y$ vs. $z$
projection).

The SciBar detector~\cite{Nitta:2004nt} was positioned upstream of the
other sub-detectors.  It consists of 14,336 extruded plastic
scintillator strips. 
Each strip has a dimension of
1.3 $\times$ 2.5 $\times$ 300~cm$^3$.  The scintillators are arranged
vertically and horizontally to construct a 3 $\times$ 3 $\times$
1.7~m$^3$ volume with a total mass of 15 tons.  
The dominant component of the SciBar detector is polystyrene ($\rm C_8H_8$).
We measured the density of the scintillator by sampling 10\% of the strips 
before the installation \cite{Maesaka:2005aj}.
The uncertainty of the total detector mass is estimated to be 1\%, 
including the effect of epoxy resin used to glue the strips.

Each strip is read
out by a wavelength shifting (WLS) fiber attached to a 64-channel
multi-anode photomultiplier (MA-PMT).  Charge information is recorded
for each channel, while timing information is recorded in groups of
32 channels by taking the logical OR with multi-hit TDC
modules~\cite{Yoshida:2004mh}.
The timing resolution for minimum-ionizing
particles, evaluated with cosmic ray data, is 1.6~ns.  The
average light yield for minimum-ionizing particles is approximately 20
photo-electrons per 1.3~cm path length, and the typical pedestal width
is below 0.3 photo-electron.  The hit finding efficiency, evaluated with
cosmic ray data, is more than 99.8\%.  The minimum length of a
reconstructable track is approximately 8~cm (three layers hit in each
view).  The track finding efficiency for single tracks of 10~cm or
longer is more than 99\%.

The MRD is located downstream of SciBar and the EC,
and is designed to measure the
momentum of muons produced by CC neutrino interactions.  It comprises
12 iron plates with thickness 5~cm sandwiched between planes of 6~mm
thick scintillation counters; there are 13 alternating horizontal and
vertical planes read out via 362 individual 2~inch PMTs.  Each iron
plate measured 274 $\times$ 305~cm$^2$.  The MRD measures the momentum
of muons up to 1.2~GeV/$c$ using the observed muon range.  Charge and
timing information from each PMT are recorded.  The average hit
finding efficiency is 99\%.

\subsection{Detector Simulation}
\label{subsec:detector_mc}


A GEANT4 framework is used for the detector simulation.  
The detector simulation includes a detailed geometric
model of the detector, including the detector frame and experimental
hall and soil.  

The Bertini cascade model within GEANT4~\cite{Heikkinen:2003sc} is used to
simulate the interactions of hadronic particles with detector
materials.  
A 10\% difference of the total pion-carbon cross section is seen for
higher energy pions between the GEANT4 simulation and external
measurements. Hence, we set $\pm10\%$ systematic uncertainty for the
pion-carbon cross section.
The systematic uncertainty of the energy deposit per unit length is
estimated to be 3\% in SciBar and MRD, and 10\% in EC, which is
dominated by the differences among various calculations of the range to
energy conversion.
Birk's constant for the SciBar scintillator was measured
to be $0.0208 \pm 0.0023 \ {\rm cm/MeV}$ \cite{Hasegawa:2006am}.  

The crosstalk of the MA-PMT is measured to be 3.15$\pm$0.4\% for
adjacent channels.  The single photo-electron resolution of the MA-PMT
is set to 50\% in the simulation, to reproduce the observed $dE/dx$
distribution of cosmic muons.  The absolute error is estimated to be
$\pm$20\%. 
A more detailed description of the detector simulation is given
in~\cite{Hiraide:2008eu}.


\section{Neutrino Interaction Simulation}
\label{sec:neutrinointeractionmc}

\subsection{Overview}
\label{ssec:neuint-overview}

We simulate neutrino interactions with carbon and hydrogen
in the SciBar detector with the NEUT
and NUANCE program libraries.
We produced separate Monte Carlo samples with NEUT and NUANCE,
and compared the two to the SciBooNE neutrino  data.

The nuclear targets handled in NEUT are protons, carbon, oxygen, and
iron.  The energy of neutrinos handled by the simulation ranges from
100~MeV to 100~TeV.  The types of neutrino interactions simulated in
both NC and CC are : elastic and quasi-elastic
scattering ($\nu N \rightarrow \ell N'$), single meson production
($\nu N \rightarrow \ell N'm$), single gamma production ($\nu N
\rightarrow \ell N' \gamma$), coherent $\pi$ production ($\nu^{12}{\rm
C (or } ^{56}{\rm Fe}) \rightarrow \ell \pi\ ^{12}{\rm C (or }
^{56}{\rm Fe})$), and deep inelastic scattering ($\nu N \rightarrow
\ell N'hadrons$), where $N$ and $N'$ are the nucleons (proton or
neutron), $\ell$ is the lepton (electron, muon or neutrino), and $m$
is the meson. In nuclei, interactions of the mesons and hadrons with
the nuclear medium are simulated following the neutrino interactions.

The types and models of neutrino interactions in NUANCE
are similar to those
in NEUT but with different cross section parameter settings in some cases
and a completely independent treatment of meson and hadron re-interactions in the 
nuclear medium.

In addition to the neutrino interactions inside SciBar,
we also simulate interactions in the EC/MRD and the surrounding 
materials (the walls of the detector hall and soil) using NEUT.
We assign a 20\% normalization uncertainty to the interaction cross sections
for both the EC/MRD and surrounding material relative to the predictions for SciBar.

\subsection{Neutrino Interaction Model}
\label{ssec:neuintmodel}

\subsubsection{Quasi-elastic scattering}
\label{sssec:qesiml}

The dominant interaction in the SciBooNE neutrino energy range 
is CC-QE scattering, which is implemented using the Smith and
Moniz model~\cite{Smith:1972xh}. 
The nucleons are treated as quasi-free particles and the
Fermi motion of nucleons along with the Pauli exclusion principle is
taken into account.  The Fermi surface momentum ($p_F$) for carbon 
is set to 217(220)~MeV/$c$
and the nuclear potential ($E_B$) is set to 25(34)~MeV/$c$
in NEUT(NUANCE), as extracted from electron scattering data \cite{Moniz:1971mt}. 
The default binding energy in NUANCE is somewhat higher because it additionally 
accounts for neutrino vs. electron scattering differences \cite{AguilarArevalo:2008fb}.
The systematic errors for $p_F$ and $E_B$ are set to $\pm$~30~MeV/$c$ and 
$\pm$~9~MeV/$c$, respectively, for both NEUT and NUANCE.

For the vector form factor, NEUT uses a dipole form with a vector mass 
of 0.84~$\rm GeV/c^2$, while NUANCE uses the BBA-2003 
form factor \cite{Budd:2003wb}.
A dipole form is used for  the axial form factor with an adjustable axial 
mass, $M_A^{QE}$, for both NEUT and NUANCE.
In NUANCE, an empirical Pauli-blocking parameter, $\kappa$,
is introduced~\cite{AguilarArevalo:2010zc} to 
better describe the MiniBooNE quasi-elastic data at low momentum transfer. 
When $\kappa > 1$, the phase space of nucleons susceptible to Pauli-blocking
is increased and hence the cross section at low momentum transfer 
is suppressed.

The values of $M_A^{QE} =1.21$~$\rm GeV/c^2$ and $\kappa = 1.000$ (i.e. no additional 
Pauli blocking adjustment) are used in NEUT, and 
$M_A^{QE} = 1.23$ $\rm GeV/c^2$ and $\kappa = 1.022$ 
are used in NUANCE \cite{AguilarArevalo:2008fb}.
A systematic uncertainty of $\pm$~0.22~GeV is assigned
 to $M_A^{QE}$ to span the difference between the value used and the global fit from 
 previous measurements \cite{Bodek:2007vi}.
The difference between $\kappa = 1.000$ and $\kappa = 1.022$ 
is also assigned as systematic uncertainty.

The same Fermi momentum distribution, nuclear potential 
are used in all other 
neutrino-nucleus interactions except for coherent
$\pi$ production.


\subsubsection{Meson production via baryon resonances}
\label{sssec:spisim}
The second most frequent interaction in SciBooNE 
is the resonant production of single pion, kaon, and eta mesons as
 described by the model of Rein and Sehgal (RS)~\cite{Rein:1980wg}.  

The RS model assumes an intermediate baryon resonance,
$N^*$, in the reaction of $\nu N \rightarrow \ell N^*, N^* \rightarrow
N'm$.  All intermediate baryon resonances with mass less than
2~GeV/$c^2$ are included.  Baryon resonances with mass greater than
2~GeV/$c^2$ are simulated as deep inelastic scattering.  
$\Delta$ re-interactions ($\Delta N \to NN$) which do 
not lead to a mesonic final state are also simulated. This re-interaction 
probability is assumed to be 0.2~$\pm$~0.2 for all $\Delta$ resonances in NEUT and 
0.1~$\pm$~0.1 (0.2~$\pm$~0.2) for $\Delta^{++/-}$ ( $\Delta^{+/0}$) resonances in NUANCE.

To determine the angular distribution of final state pions, 
the RS method~\cite{Rein:1987cb} is used for the $P_{33}(1232)$
resonance in both NEUT and NUANCE. 
For other resonances, the directional distribution of the
generated pion is chosen to be isotropic in the resonance rest frame.

The axial-vector form factors are formalized to be dipole with
$M_A^{1\pi}=1.21$~$\rm GeV/c^2$ for NEUT and $M_A^{1\pi}=1.10$~$\rm GeV/c^2$
for NUANCE, with an uncertainty of 
0.28~$\rm GeV/c^2$ in both cases.

An additional uncertainty is assigned to account for the observed
$Q^2$ disagreement between the SciBooNE CC 1$\pi$-enriched data
samples and NEUT \cite{Hiraide:2008eu}.  A similar disagreement is
also observed for NUANCE. The size of the uncertainty is determined by
re-weighting CC resonant pion events as a function of true $Q^2$ such
that they match the observed distribution in the SciBooNE data, and
using the difference between the re-weighted distribution and the
central value as the uncertainty.

Resonance decays leading to multi-pion final states are also included
in the model and are simulated assuming $M_A^{N\pi}=1.30 \pm 0.52$~GeV/$c^2$ 
in NUANCE. This value of $M_A^{N\pi}$ is chosen strictly to
ensure that the total CC cross section prediction reproduces previous
experimental data.
In NEUT, multi-pion
production is simulated as deep inelastic scattering as described
later in Sec.~\ref{sssec:delsim}, and the RS model is
not used.  The size of the systematic uncertainty is estimated based on
$M_A^{N\pi}$ variations for NUANCE, and the same size error is
assumed also for the NEUT prediction.

\subsubsection{Coherent pion production}
\label{sssec:cohpisim}
Coherent pion production is a neutrino interaction with a nucleus
which remains intact, releasing one pion with the same charge as the incoming 
weak current.
Because of the small momentum transfer 
to the target nucleus, the outgoing pion tends to be emitted in 
the forward direction, closely following the incoming neutrino direction.
The formalism developed by Rein and
Sehgal~\cite{Rein:1982pf,Rein:2006di} is used to simulate such
interactions.  The axial vector mass, $M_A^{coh}$, is set to 
$1.0 \pm 0.28$~GeV/$c^2$ in both NEUT and NUANCE.
In NEUT, the total and inelastic pion-nucleon cross sections
 from the original Rein-Sehgal publication are employed~\cite{Rein:1982pf,Rein:2006di}. 
In NUANCE, they are obtained from fits to PDG data~\cite{Nakamura:2010zzi}
and implemented as a function of pion energy.
Additionally, the NC and CC coherent pion production cross section 
predictions in NUANCE are rescaled by a factor of 0.65 to better 
match the measured rate of NC coherent $\pi^0$ production 
as measured in MiniBooNE~\cite{AguilarArevalo:2008xs}.

\subsubsection{Deep inelastic scattering}
\label{sssec:delsim}
The deep inelastic scattering (DIS) cross section is calculated using
the GRV98 parton distribution functions~\cite{Gluck:1998xa}.
Additionally, we have included the corrections in the small $Q^2$ region 
developed by Bodek and Yang\cite{Bodek:2003wd} for both NEUT and NUANCE.
The implementation of the 
model is slightly different in NEUT and NUANCE. 

In NEUT, the DIS contribution 
is included for hadronic invariant masses $W >1.3$~GeV/$c^2$. The pion multiplicity 
is additionally restricted to be greater than one for $1.3<W<2$~GeV/$c^2$ 
to avoid double-counting sources of single pion production that are already included 
in the resonance portion of the simulation. 
The multi-hadron final states are simulated with two models in NEUT: a 
custom-made program \cite{Nakahata:1986zp} for events with W between 
1.3 and 2.0~GeV/$c^2$ and PYTHIA/JETSET \cite{Sjostrand:1993yb} for events 
with W larger than 2~GeV/$c^2$.

A restriction on pion multiplicity is not enforced by NUANCE.
Instead, the DIS contribution slowly increases for $W$ values starting at 1.7~GeV 
and becomes the only source of neutrino interactions above $W>2$~GeV. This is done 
to create a smooth transition between the resonance and DIS models and ensure
continuity in distributions of kinematics and hadron multiplicity in the region of 
overlap.

\subsubsection{Intra-nuclear interactions}
\label{sssec:fsisim}

Following production, the intra-nuclear interactions of mesons
and nucleons are simulated using a cascade model in which the
particles are traced until they escape from the nucleus. 

Although we only use kinematic information from the final state muon
in this analysis, the simulation of intra-nuclear interactions is
important since the pions/protons emitted from the nucleus can be
mis-reconstructed as muons.

The inelastic scattering, charge exchange and absorption of pions in
nuclei are simulated. 
For inelastic scattering and charge
exchange interactions, the direction and momentum of pions are
affected.  In the scattering amplitude, Pauli blocking is also taken
into account.
A more detailed description of the intra-nuclear interaction simulations
in NUANCE and NEUT can be found elsewhere~\cite{Casper:2002sd,Hiraide:2008eu}.

A 25\% (30\%) uncertainty in the overall pion absorption (charge
exchange) cross section is assumed based on comparisons to pion-carbon scattering
data \cite{Ashery:1981tq, *Ransome:1992bx, *Jones:1993ps}. The 
uncertainty in proton re-scattering is estimated to be 10\%. Hence, we apply a 10\% 
error on the number of proton tracks observed in SciBar.
Additionally, we set a 20\% uncertainty on the NC/CC ratio, 
estimated from the model dependence of the lepton-mass effect 
in the small $Q^2$ region.

\subsection{The Expected Number of Neutrino Events}

Table~\ref{tab:xsec_params} summarizes the parameter choices used in NEUT and NUANCE for
the comparisons presented here. 
We chose these parameters since they are the default parameter settings 
used in Super-K, K2K and T2K (NEUT) and MiniBooNE (NUANCE).
The different parameter values result in different
neutrino cross section predictions between the two.

\begin{table}[htbp]
  \centering
  \caption{Parameters used for neutrino interaction simulation.}
  \label{tab:xsec_params}
  \begin{ruledtabular}
  \begin{tabular}{cccc}
    Parameter  & NEUT & NUANCE  \\
    \hline
    $p_{F}$   & 217 MeV &  220 MeV\\
    $E_{B}$   & 25 MeV  &  34 MeV \\
    $M_A^{QE}$     & 1.21 GeV  &  1.23 GeV\\
    $\kappa$  & 1.00      &  1.022    \\
    $M_A^{1\pi}$ & 1.21  GeV &1.10 GeV\\
    $M_A^{coh}$ & 1.0  GeV &1.0 GeV\\
    $M_A^{N\pi}$ & (DIS) &1.3 GeV\\
  \end{tabular}
  \end{ruledtabular}
\end{table}

With the SciBooNE neutrino beam exposure of $0.99 \times 10^{20}$
protons on target, the expected number of events in the SciBooNE
detector for each neutrino interaction is listed in
Table~\ref{ta:neut-nuance}.  Because of the difference in parameter
choices (Table~\ref{tab:xsec_params}), NEUT predicts a larger QE and
single pion rate than NUANCE. The difference in QE rate is largely
coming from the choice of $\kappa$ values, and the difference in
single pion rate can be largely accounted for by the difference in
$M_A^{1\pi}$ assumptions.  The major source of the factor of two
larger multi-pion/DIS rate in NEUT compared to NUANCE is the
difference of multi pion production simulation in the range $1.3 < W <
2.0$~GeV; NEUT simulate these events as DIS, while NUANCE uses a
resonant production model.  These differences in cross section
predictions between similar models with perfectly reasonable parameter
choices further highlight the inherent uncertainty in neutrino
generator predictions and stress the importance of additional neutrino
interaction measurements in this region.

\begin{table*}[htbp]
\caption{ The expected number and fraction of events in each neutrino
  interaction estimated by NEUT and NUANCE at the SciBooNE detector
  location with the neutrino beam exposure of $0.99 \times 10^{20}$
  protons on target.  The 10.6~ton fiducial volume of the SciBar
  detector is assumed.  CC and NC interactions are abbreviated as CC
  and NC, respectively.}
\label{ta:neut-nuance}
\begin{center}
  \begin{ruledtabular}
 \begin{tabular}{lrrrr}
 
                                       &  \multicolumn{2}{c}{NEUT} &  \multicolumn{2}{c}{NUANCE} \\
    Interaction Type                   &  \# Events & Fraction(\%) &  \# Events & Fraction(\%) \\
   \hline                                                                                      
    CC quasi-elastic                   &  53,038  &  41.5          &  47,573  &  43.6          \\
    CC single $\pi$ via resonances     &  29,452  &  23.0          &  25,863  &  23.7          \\
    CC coherent $\pi$                  &   1,760  &   1.4          &   1,736  &   1.6          \\
    CC multi-pion, DIS, etc             &   6,834  &   5.3          &   3,140  &   2.9          \\
    NC total                           &  36,836  &  28.8          &  30,734  &  28.2          \\
    \hline
    Total  &  127,920  &  100.0  &  109,046  &  100.0 \\
 \end{tabular}
 \end{ruledtabular}
\end{center}
\end{table*}

\begin{figure}[htbp]
  \centering
  \includegraphics[width = \columnwidth]{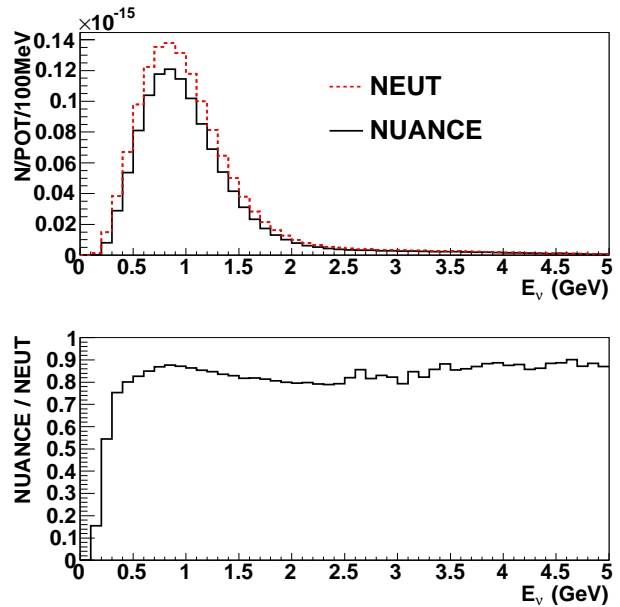}
  \caption{(color online).
    The neutrino energy distributions of $\nu_\mu$ CC interactions 
    at the SciBar detector
    generated by NEUT (red dashed line) and NUANCE (black solid line).
    Top plot is the number of interactions per unit POT,
    and the bottom plot shows the 
    ratio between the NEUT and NUANCE predictions.
  }
  \label{fig:neut-nuance-enu}

\end{figure}

Figure~\ref{fig:neut-nuance-enu} shows the expected number of total
$\nu_\mu$ CC interactions as a function of neutrino energy. One can
see that the cross section prediction from NEUT is about 10 - 20\%
larger than that from NUANCE across the range of SciBooNE energies.

The expected number of  $\nu_\mu$ CC interactions
in the 10.6~ton SciBar fiducial volume (Sec.~\ref{subsec:event_selection}) 
are $9.11 \times 10^4$ and $7.83 \times 10^4$
for the NEUT and NUANCE predictions, respectively.


\section{ CC Event Reconstruction and Selection}
\label{sec:reconstruction}

To measure the rate of CC inclusive interactions,
we use neutrino events occurring in SciBar 
with a muon in the final state.
We select muons originating in the SciBar fiducial volume (FV),
defined to be 
$\pm 130$~cm in both the $x$ and $y$ dimensions, and $2.62 < 
z < 157.2$~cm, a total mass of 10.6~tons.

We describe the reconstruction of muon tracks 
in Sec.~\ref{subsec:track-reconstruction},
the event selections and details the of sub-samples 
in Sec.~\ref{subsec:event_selection}
and comparison of the data to the MC predictions 
in Sec.~\ref{subsec:basic_distribution}.

\subsection{Track Reconstruction}
\label{subsec:track-reconstruction}

\subsubsection{Track finding}
\label{subsec:track-finding}

The first step of the event reconstruction is to search for 
two-dimensional tracks in each view of SciBar using a cellular
automaton algorithm~\cite{Alcaraz-thesis}.  
Three dimensional tracks are reconstructed by matching the
timing and $z$-edges of the two dimensional projections;
differences between two two-dimensional projections are required to be
less than 50~ns, and the $z$-edge difference must be less than 6.6~cm
for both upstream and downstream edges. 
The two dimensional tracks in MRD are  independently reconstructed 
using hits in the MRD clustered within a 50~ns timing window.
Three dimensional tracks in the MRD are reconstructed by matching the
timing of the two dimensional projections.

Then, if the downstream edge of a SciBar track lies in last two layers
of SciBar, we search for a track or hits in the MRD that are matched
with the SciBar track.  For matching an MRD track to a SciBar track,
the upstream edge of the MRD track is required to be on either one of
the first two layers of the MRD, and to be within 30~cm of the
projected entry point of the SciBar track into the MRD in each view. A
SciBar track matched with MRD is defined as a SciBar-MRD matched
track.  The matching criteria impose a muon momentum threshold of
350~MeV/$c$.  A more detailed description of the track reconstruction
can be found elsewhere\cite{Hiraide:2008eu}.

\subsubsection{Track Classification}
\label{subsec:track-types}

\begin{figure}[htbp]
  \centering
  \includegraphics[width = \columnwidth]{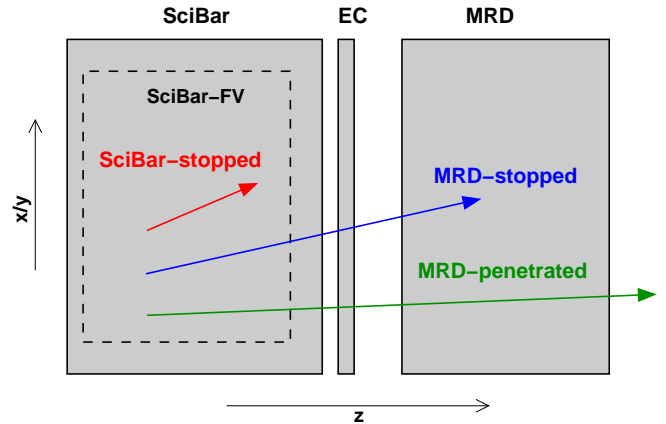}
  \caption{
    (color online). Types of tracks used for this CC 
    interaction measurement.
  }
  \label{fig:cc-inclusive-samples}
\end{figure}

We define three types of tracks used in this analysis: 
SciBar-stopped, MRD-stopped and MRD-penetrated tracks,
as shown in Fig.~\ref{fig:cc-inclusive-samples}.

Tracks with both edges contained in the SciBar FV are 
classified as SciBar-stopped tracks.
MRD-stopped and MRD-penetrated tracks are both 
subsets of the SciBar-MRD matched sample.
An MRD-stopped track is selected by requiring 
the downstream edge of the track to be within
$|x| < 132$~cm , 
$|y| < 111$~cm , and
$z < $ (last layer of the MRD).
An MRD-penetrated track is selected by requiring  
additional hits at the most downstream scintillator of the  MRD.
Tracks which exit  from the side of the MRD are not used in this analysis.

\subsubsection{Particle Identification}
\label{subsec:particel_id}

The SciBar detector has the capability to 
distinguish protons from other particles using $dE/dx$.
We define a muon~confidence~level (MuCL)
using the observed energy deposit per layer for all reconstructed 
tracks~\cite{Hiraide:2008eu}. 
Tracks with MuCL greater than 0.05 are considered muon-like (or pion-like)
and the others are classified as proton-like.

\subsubsection{Muon Kinematics Reconstruction}
\label{subsec:muon-kinematics-reconstructoin}

The slopes of the muon angles with respect to the beam in the 
two SciBar views are used to calculate the three dimensional muon 
angle with respect to the beam ($\theta_\mu$).

According to the MC simulation, 
about 30 \% of SciBar-stopped tracks are expected to be backward-going. 
To identify these backward-going tracks, we use the 
delayed timing signal produced by the decay electrons 
from stopped muons.
A track is identified to be backward-going if it has
(1) at least one delayed hit at $t > 200$~nsec at the upstream ends 
of the two dimensional views, and  
(2) no delayed hit at $t > 200$~nsec at the downstream ends.
We impose this requirement for both top- and side- PMT signals from
the track endpoints to remove hits due to random noise and
after pulsing  from the PMTs.
The performance of this identification is estimated using a MC sample
of CCQE events with 1 reconstructed track.
In this sample, 
the efficiency of selecting backward-going tracks is 
$\sim$ 57\%,
and the probability of mis-reconstructing a forward track as
backward is $\sim$1\%.
The loss of efficiency is predominantly  due to
decay electrons emitted 
at a large angle, producing hits in either the top or side PMTs,
but not both.

This track direction identification is only applied to
SciBar-stopped tracks.
All tracks reaching the MRD are assumed to be forward-going,
because the fraction of backward tracks is estimated to be small 
 ($\sim$ 4\%) and also 
a similar tagging of decay electrons is not possible in the MRD
since the electrons stop and are undetected within the steel plates.

The kinetic energy of the muon ($E_{kin}$) is calculated by the range and the expected
energy deposition per unit length ($dE/dx$) in the detector materials
\begin{equation}
E_{kin} =  E^{SciBar} +  E^{EC}  +  E^{Wall}  +  E^{MRD},
\end{equation}
where $E^{SciBar}$, $E^{EC}$, $E^{Wall}$ and $E^{MRD}$ are the
expected energy deposit by muons in SciBar, the EC, the wall of the
dark box between the EC and MRD, and the MRD, respectively.

For SciBar-stopped tracks, $E^{EC},~E^{Wall}$ and $E^{MRD}$
are set to 0, and $E^{SciBar}$ is calculated by a 
range to energy look-up table based on the MC simulation.

For MRD-stopped  tracks, 
energy deposits in SciBar, EC and the wall are computed as
$E^{SciBar} = 2.04~{\rm MeV/cm} \times L_{SB}~(\rm cm)$, 
$E^{EC} = 90.8/\cos \theta_\mu ~\rm{MeV}$ and 
$E^{Wall}= 3.3/\cos \theta_\mu ~\rm{MeV}$,
where $L_{SB}$ is the reconstructed track length in SciBar.
$E^{MRD}$ is calculated by a 
range to energy look-up table based on the MC simulation.

The average muon angular resolution is $0.9^{\circ}$ for all samples.
The muon momentum resolutions are 
15~MeV/c for SciBar-stopped and 50~MeV/c for MRD-stopped tracks, respectively.

\subsection{Event Selection and Classification}
\label{subsec:event_selection}

\subsubsection{Event Selection}

We select the highest momentum track  with MuCL~$>$~0.05
in an event
as a muon candidate.
We also require the reconstructed momentum to be greater than
0.25~GeV/c to reject short proton or pion tracks from 
neutral current interactions.
Then, we require the upstream edge of the muon candidate to be in 
the SciBar FV.

\subsubsection{Event Classification}
\label{subsec:event-classification}

Events with muon track candidates are subdivided into
three sub-samples: 
SciBar-stopped, MRD-stopped and 
MRD-penetrated samples,
according to the track classification given 
in Sec.~\ref{subsec:track-types} for the muon candidate.

\begin{figure}[htbp]
  \centering
  \includegraphics[width = \columnwidth]{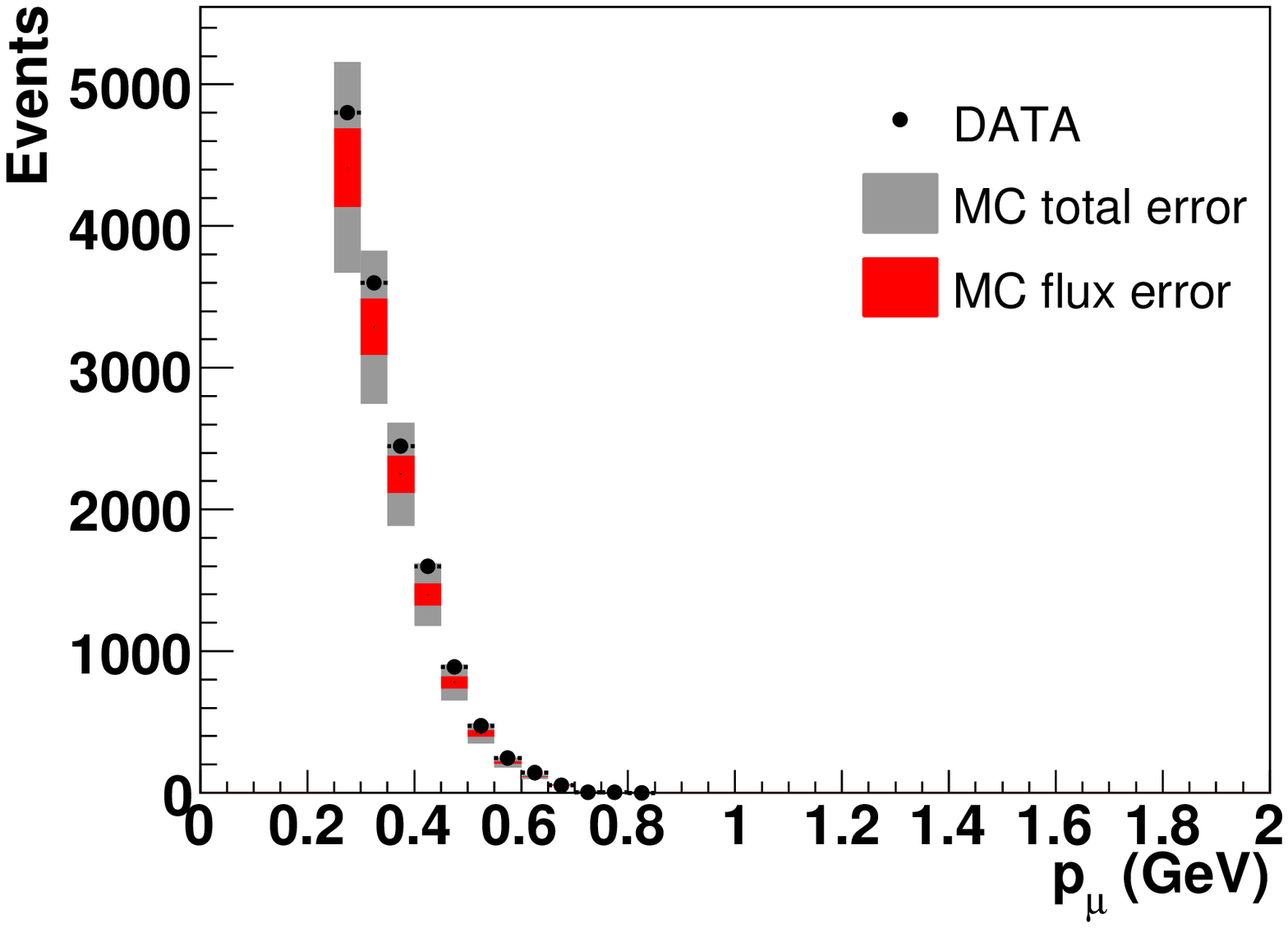}
  \includegraphics[width = \columnwidth]{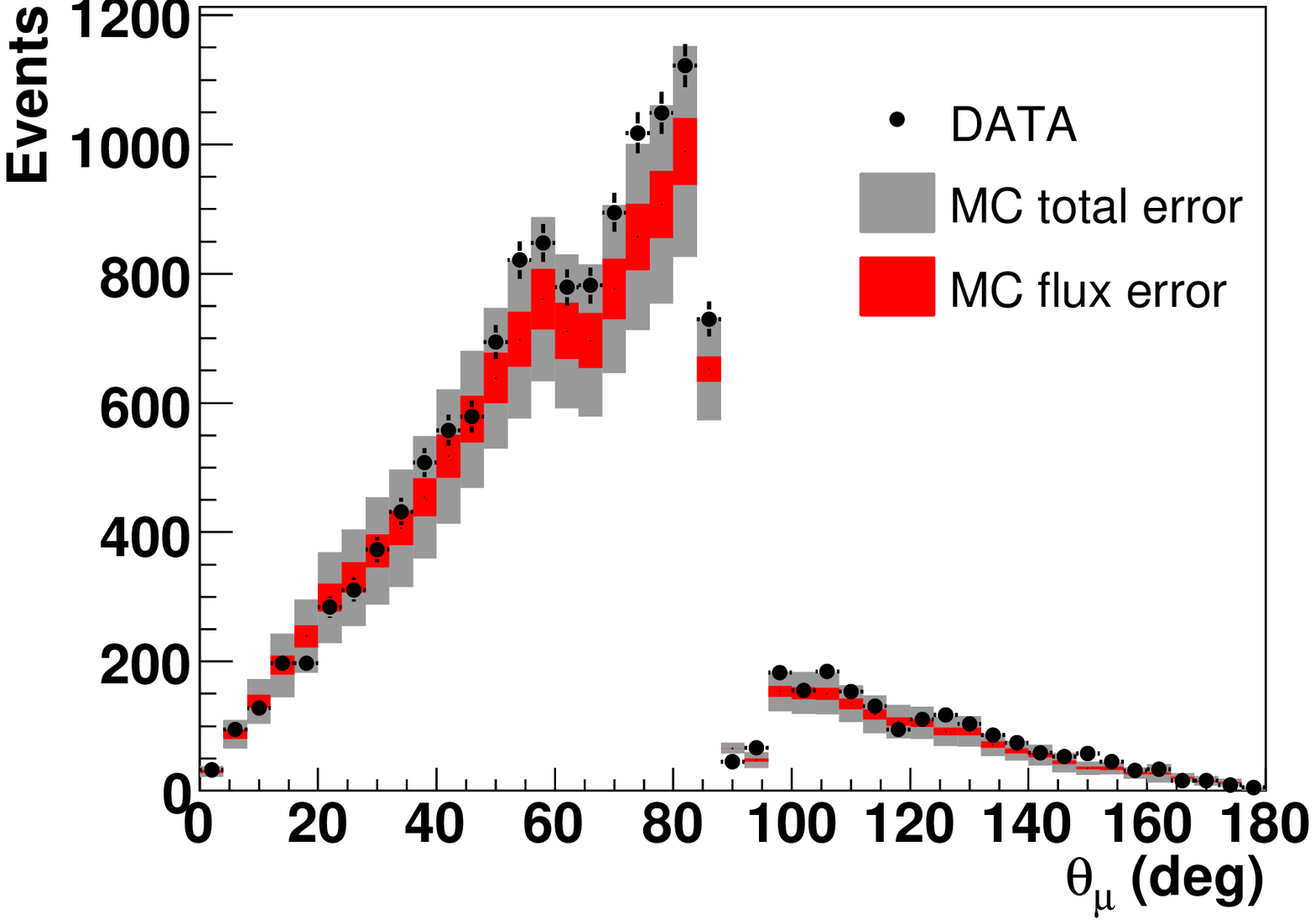}
  \caption{(color online).
    Distributions of reconstructed momentum (top) and 
    angle (bottom) of the muon candidate in the SciBar-stopped sample.
    The MC prediction is based on NEUT and absolutely normalized by the 
    number of POT.
    The total and flux systematic errors on the MC predictions are separately shown.
    The dominant source of the total error is the cross section uncertainty.
  }
  \label{fig:pmu-thetamu-sb}
\end{figure}

\paragraph{SciBar-stopped sample}
The SciBar-stopped sample provides the lowest energy sample;
the mean energy of neutrinos in this sample is 1.0~GeV. 
According to the simulation, the purity of $\nu_\mu$ CC interactions 
in this sample is 85\%. Impurities are due to
$\nu_\mu$ NC interactions ($\sim$~7\%),
interactions occurring in  the surrounding material ($\sim$ 5\%) and
$\overline{\nu}_\mu$ CC interactions ($\sim$ 1.5\%). 
Figure~\ref{fig:pmu-thetamu-sb} shows the distributions of the reconstructed muon 
momentum ($p_\mu$) and angle ($\theta_\mu$) for SciBar-stopped muons. 
The backward-going muon identification described in Sec.~\ref{subsec:muon-kinematics-reconstructoin}
is used for this sample. 
All tracks which are not identified as backward-going 
are accounted as forward-going tracks for both data and MC samples.
The expected number of events in each interaction mode is summarized in
Table~\ref{ta:neut-nuance-sbstop}.
The contamination of the cosmic-ray backgrounds is estimated,
using off-beam data, to be $671.2 \pm 11.6$ events,
where the uncertainty comes from the statistics of the off-beam data sample.
This rather large contamination ($\sim 5\%$) is due to 
cosmic-rays penetrate SciBar vertically, which leave sparse hits and are often 
mis-reconstructed as short tracks.
This effect is negligible for the other sub-samples.

\begin{table}[htbp]
\caption{
 The expected number and fraction of events  in each neutrino interaction type
for the SciBar-stopped sample,
as estimated by NEUT and NUANCE.
The external events comprise interactions 
in the EC, MRD and surrounding material.
}
\label{ta:neut-nuance-sbstop}
\begin{center}
 \begin{ruledtabular}
 \begin{tabular}{lrrrr}
    Interaction           &  \multicolumn{2}{c}{NEUT} &  \multicolumn{2}{c}{NUANCE} \\
     type                   &  Events & Fraction(\%) &  Events & Fraction(\%) \\
    \hline
    CC QE          &  5818 & 47.4  & 5511 & 50.8\\                                                                             
    CC res. 1$\pi$ &  3694 & 30.1  & 3226 & 29.8\\
    CC coh. 1$\pi$ &  119  & 1.0   & 123  & 1.1\\
    CC other       &  914  & 7.4   & 350  & 3.2\\
    NC             &  916  & 7.4   & 842  & 7.8\\
    All non-$\nu_\mu$ &  188  & 1.5   & 161  & 1.5\\
    External       &  629  & 5.1   & 629  & 5.8\\
    \hline
      Total  &  12278  &   &  10,842  &  \\
 \end{tabular}
 \end{ruledtabular}
\end{center}
\end{table}

\paragraph{MRD-stopped sample}

\begin{figure}[htbp]
  \centering
  \includegraphics[width = \columnwidth]{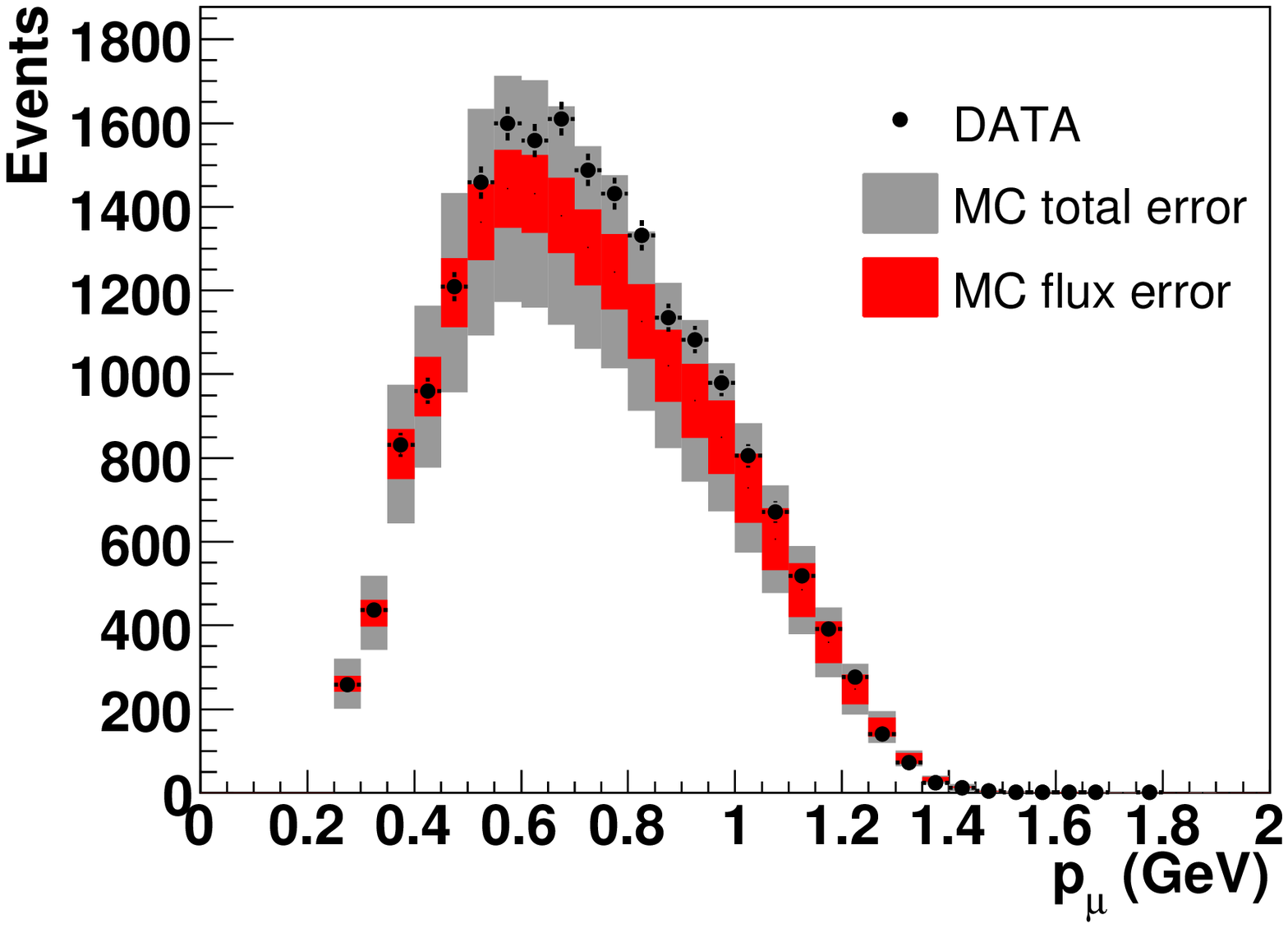}
  \includegraphics[width = \columnwidth]{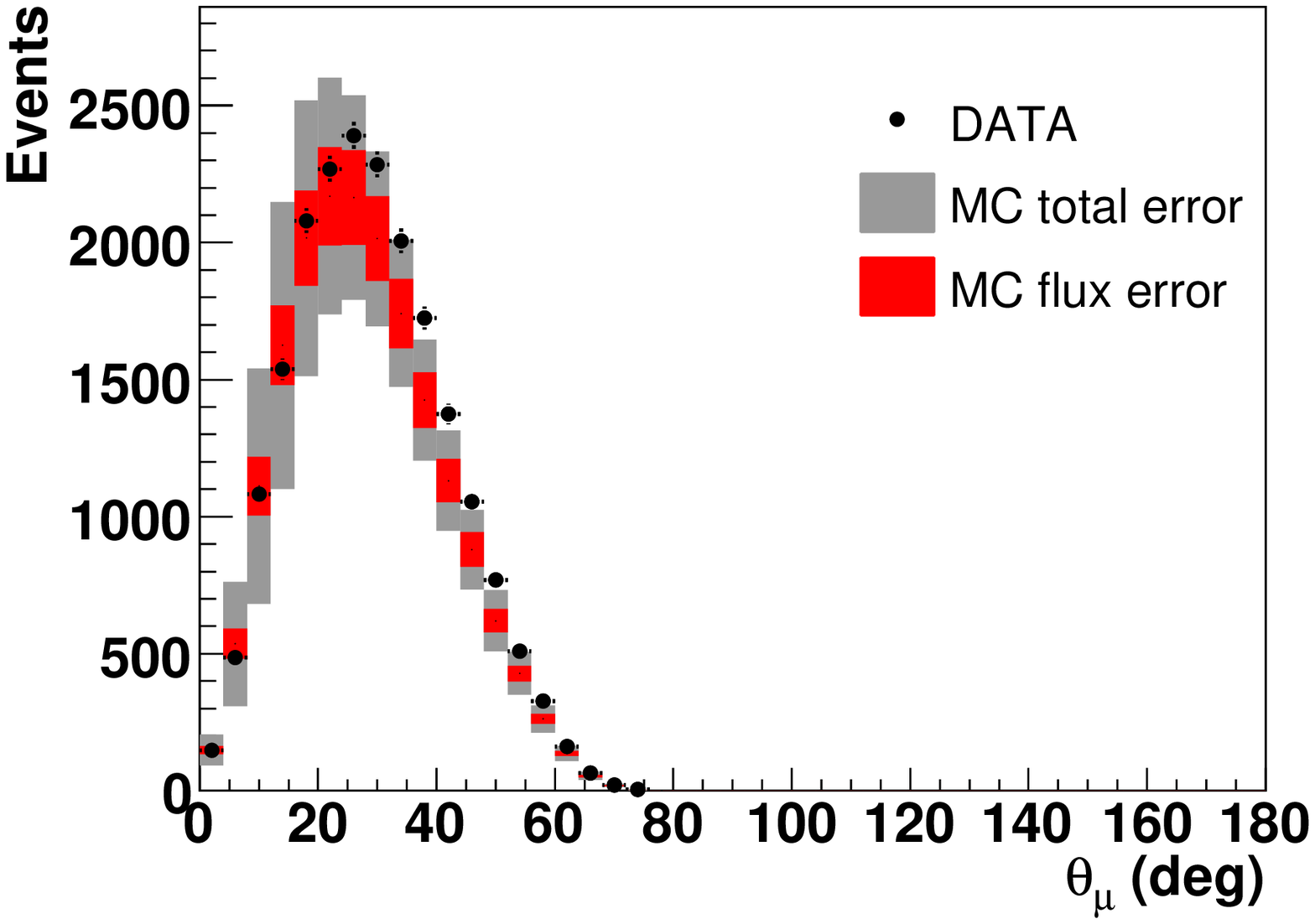}
  \caption{(color online).
    Distributions of reconstructed momentum (top) and 
    angle (bottom) of the muon candidate in the MRD-stopped sample.
    The MC prediction is based on NEUT and absolutely normalized by the 
    number of POT.
    The total and flux systematic errors on the MC predictions are separately shown.
    The dominant source of the total error is the cross section uncertainty.
  }
  \label{fig:pmu-thetamu-mrd}
\end{figure}

The MRD-stopped sample has the largest statistics among the three.
The mean energy of neutrinos in the MRD-stopped sample is
1.2~GeV.
According to the simulation, the purity of $\nu_\mu$ CC interaction 
in this sample is 91\%. Impurities are from
neutrino interactions in the EC/MRD which back-scatter ($\sim$~4\%),
$\nu_\mu$ NC interactions ($\sim$ 3\%)
and $\overline{\nu}_\mu$ CC interactions ($\sim$ 2\%). 
Figure~\ref{fig:pmu-thetamu-mrd} shows the distributions of the reconstructed muon 
momentum ($p_\mu$) and angle ($\theta_\mu$) for MRD-stopped muons. 
The expected number of events in each interaction mode is summarized in
Table~\ref{ta:neut-nuance-mrdstop}.
The contamination of the cosmic-ray backgrounds is estimated 
to be $54.6 \pm 3.3$ events.

\begin{table}[htbp]
\caption{
 The expected number and fraction of events  in each neutrino interaction type
for the MRD-stopped sample,
as estimated by NEUT and NUANCE.
The external events comprise interactions 
in the EC, MRD and surrounding material.}
\label{ta:neut-nuance-mrdstop}
\begin{center}
 \begin{ruledtabular}
 \begin{tabular}{lrrrr}
    Interaction           &  \multicolumn{2}{c}{NEUT} &  \multicolumn{2}{c}{NUANCE} \\
     type                   &  Events & Fraction(\%) &  Events & Fraction(\%) \\
    \hline
    CC QE               & 10341 & 56.1 & 8385 & 52.3  \\
    CC res. 1$\pi$      & 4789 & 26.0  & 4839 & 30.2 \\
    CC coh. 1$\pi$      & 659 & 3.6  & 633 & 3.9   \\
    CC other            & 1010 & 5.5   & 600 & 3.7 \\
    NC                  & 577 & 3.1  & 569 & 3.5  \\
    All non-$\nu_\mu$   & 320 & 1.7  & 281 & 1.8  \\
    External            & 729 & 4.0  & 729 & 4.5  \\
    \hline
      Total  &  18,427  &    &  16,036  &  \\
 \end{tabular}
 \end{ruledtabular}
\end{center}
\end{table}

\begin{figure}[htbp]
  \centering
  \includegraphics[width = \columnwidth]{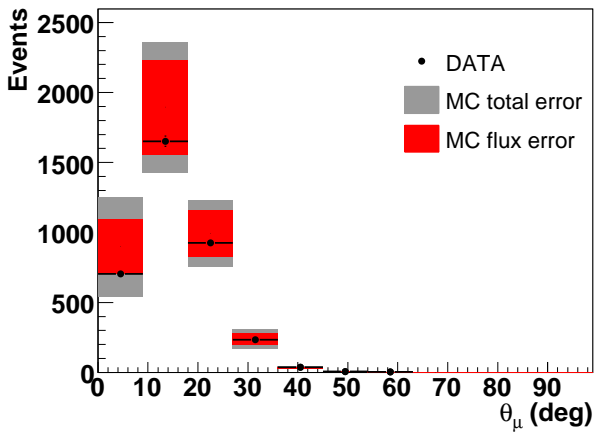}
  \caption{(color online).
    Distribution of reconstructed angle of the muon candidate 
    in the MRD-penetrated sample.
    The MC prediction is based on NEUT and absolutely normalized by the 
    number of POT.
    The total and flux systematic errors on the MC predictions are separately shown.
  }
  \label{fig:pmu-thetamu-pene}
\end{figure}

\paragraph{MRD-penetrated sample}
The MRD-penetrated sample is the highest energy sample among the three.
The mean energy of neutrinos is 2.4~GeV.
Although the track angles can be measured, we do not have the capability to 
reconstruct the muon momentum for the tracks which exit the MRD.
However, this sample can provide the normalization for the 
highest energy region. 
Hence, this sample is also used for the 
neutrino interaction rate measurement.
According to the simulation, the purity of $\nu_\mu$ CC interaction 
in this sample is 97\%. Impurities mostly come from 
$\overline{\nu}_\mu$ CC interactions ($\sim$ 2\%). 
Figure~\ref{fig:pmu-thetamu-pene} shows the distributions of the reconstructed muon 
angle ($\theta_\mu$) of the MRD-penetrated muons. 
The expected number of events in each interaction mode is summarized in
Table~\ref{ta:neut-nuance-mrdpene}.
The contamination of the cosmic-ray backgrounds is estimated 
to be $16.6 \pm 1.8$ events.

\begin{table}[htbp]
\caption{
 The expected number and fraction of events  in each neutrino interaction type
for the MRD-penetrated sample,
as estimated by NEUT and NUANCE.
The external events comprise interactions 
in the EC, MRD and surrounding material.
}
\label{ta:neut-nuance-mrdpene}
\begin{center}
 \begin{ruledtabular}
 \begin{tabular}{lrrrr}
    Interaction           &  \multicolumn{2}{c}{NEUT} &  \multicolumn{2}{c}{NUANCE} \\
     type                   &  Events & Fraction(\%) &  Events & Fraction(\%) \\
    \hline
    CC QE              & 2428 & 60.0   &  1943 & 57.0 \\
    CC res. 1$\pi$     & 1008 & 24.9   &  976 & 28.6  \\
    CC coh. 1$\pi$     & 140 & 3.5     &  130 & 3.8   \\
    CC other           & 356 & 8.8   &  255 & 7.4   \\
    NC                 & 1.5 & 0.04 &  2.3 & 0.07  \\
    All non-$\nu_\mu$  & 89  & 2.2   &  75 & 2.2   \\
    External           & 27  & 0. 7 &  27 & 0.8  \\
    \hline
      Total  & 4049   &    &  3407  &   \\
 \end{tabular}
 \end{ruledtabular}
\end{center}
\end{table}

\subsubsection{Efficiency Summary}

\begin{figure}[htbp]
  \centering
  \includegraphics[width = \columnwidth]{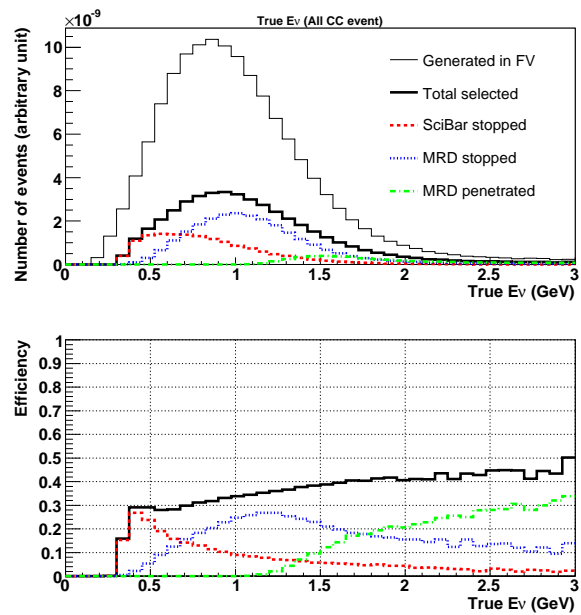}
  \caption{(color online).
    (Top) Number of CC events in the SciBar FV as a function of true  $E_\nu$, 
    predicted by the NEUT based simulation.
    The number of selected events in each sub-sample are also shown.
    (Bottom) Detection efficiency as a function of true neutrino energy for each 
    sub-sample.}
  \label{fig:efficiency_enu}
\end{figure}

Figure~\ref{fig:efficiency_enu} shows the efficiency of CC events 
as a function of 
true neutrino energy for each sub-sample, estimated
from the NEUT based MC simulation.
By combining these three samples, we can obtain fairly uniform 
acceptance for neutrinos above 0.4~GeV.

\subsection{Data Comparison to the MC prediction}
\label{subsec:basic_distribution}

Table~\ref{tab:nevents} shows the number of events obtained 
from data and the predictions from NEUT and NUANCE based
MC simulations.
The contamination of cosmic-ray backgrounds is estimated 
using the off-beam data, and have been subtracted from the data.
For the total number of events from the three sub-samples,
we find a data/MC normalization factor of
 1.08 for the NEUT prediction, and 1.23 for the NUANCE prediction.

\begin{table*}[htbp]
  \centering
  \caption{The number of events in each sub-sample from the data and 
the predictions from NEUT/NUANCE-based MC. The numbers in 
parentheses show the ratio between the data and the predictions.
The cosmic-ray backgrounds are estimated from off-timing data and subtracted
from the data.}
  \label{tab:nevents}
 \begin{ruledtabular}
  \begin{tabular}{ccccc}
  Sample  & SciBar-stopped & MRD-stopped & MRD-penetrated & Total\\
  \hline
  Data& 13588.8 &  20236.4 & 3544.4 & 37369.6 \\
  NEUT& 12278.3(1.11) & 18426.3(1.10) & 4049.0(0.88) &34753.6(1.08) \\
  NUANCE& 10841.9(1.25) & 16036.2(1.26) & 3407.5(1.04) & 30285.6(1.23) \\
  \end{tabular}
 \end{ruledtabular}
\end{table*}

To compare the MC predictions with data,
the neutrino energy($E_{\nu}$) and  the square of the four-momentum transfer($Q^2$) 
are the key  variables since  a flux variation is purely a function of  $E_{\nu}$
while a variation of the cross section model typically changes the $Q^2$
distribution.
We reconstruct these variables assuming CC-QE interaction kinematics.
The reconstructed $E_{\nu}$ is calculated as
\begin{equation}
\label{eq:pearsons-enurec}
E_{\nu}^{rec} = \frac{m_p^2 - (m_n - E_B)^2 - m_{\mu}^2 + 2(m_n - E_B)E_{\mu}}
{2(m_n - E_B - E_{\mu} + p_{\mu} \cos {\theta}_{\mu} )},
\end{equation}
where $m_p$, $m_n$ and $m_{\mu}$ are the mass of proton, neutron and muon, respectively, 
$E_\mu$ is the muon  total energy, and $E_B$ is the nuclear potential energy. 
The reconstructed $Q^2$ is given by,
\begin{equation}
\label{eq:pearsons-q2rec}
Q^2_{rec} = 2 E_{\nu}^{rec}(E_{\mu} - p_{\mu} \cos {\theta}_{\mu}) - m_{\mu}^2.
\end{equation}

\begin{figure*}[htbp]
  \centering
   \includegraphics[width = \columnwidth]{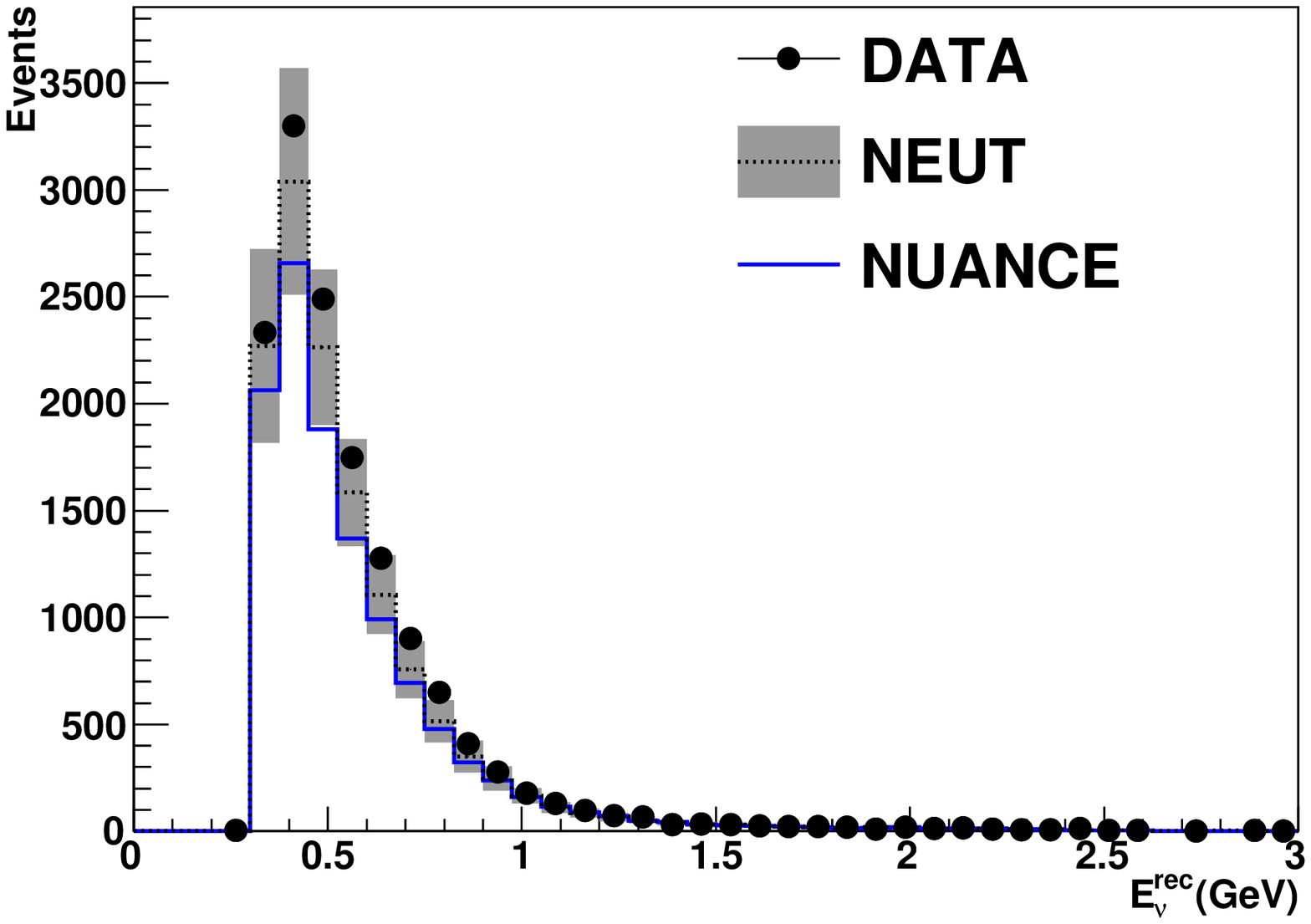}
   \includegraphics[width = \columnwidth]{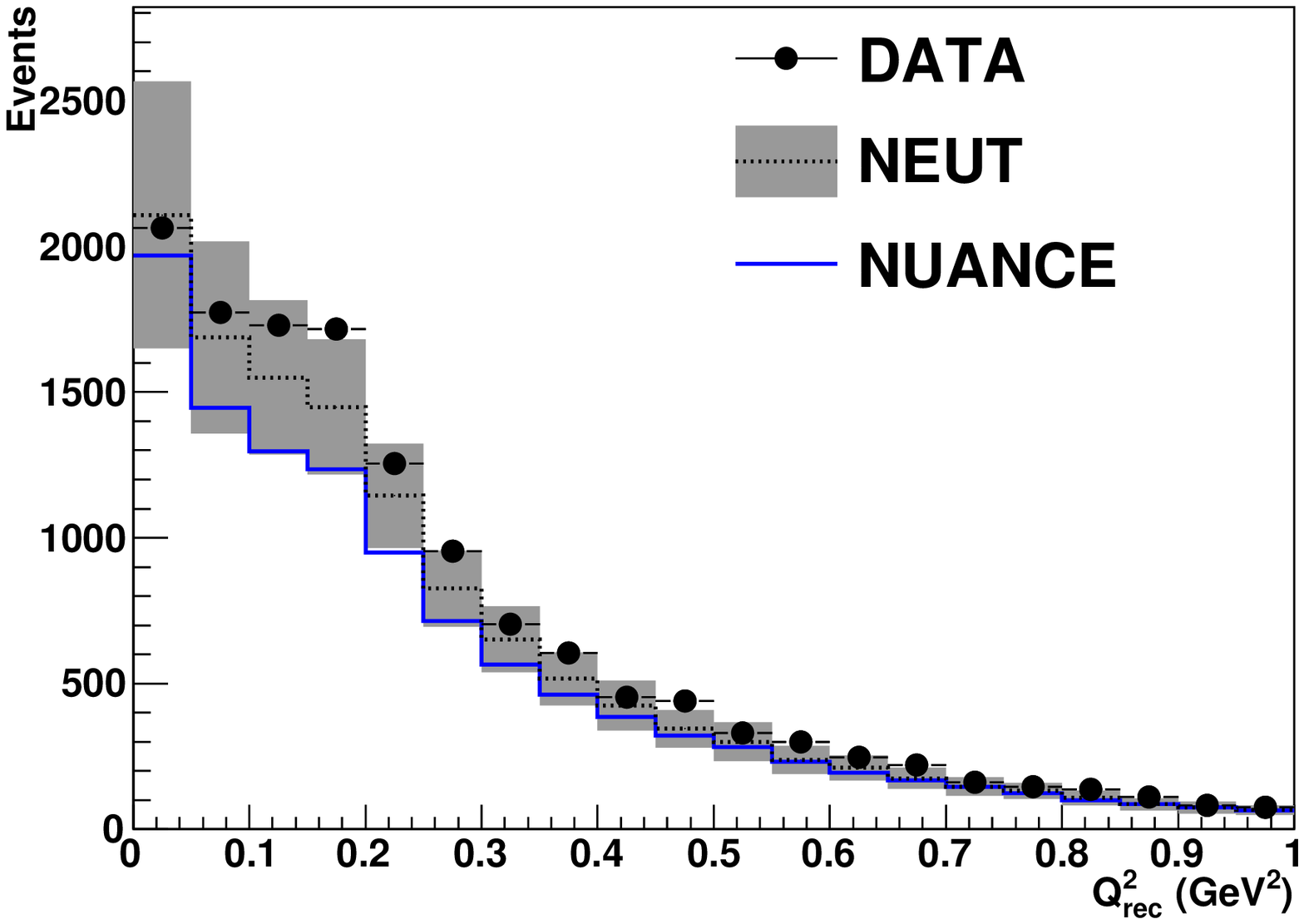}
   \includegraphics[width = \columnwidth]{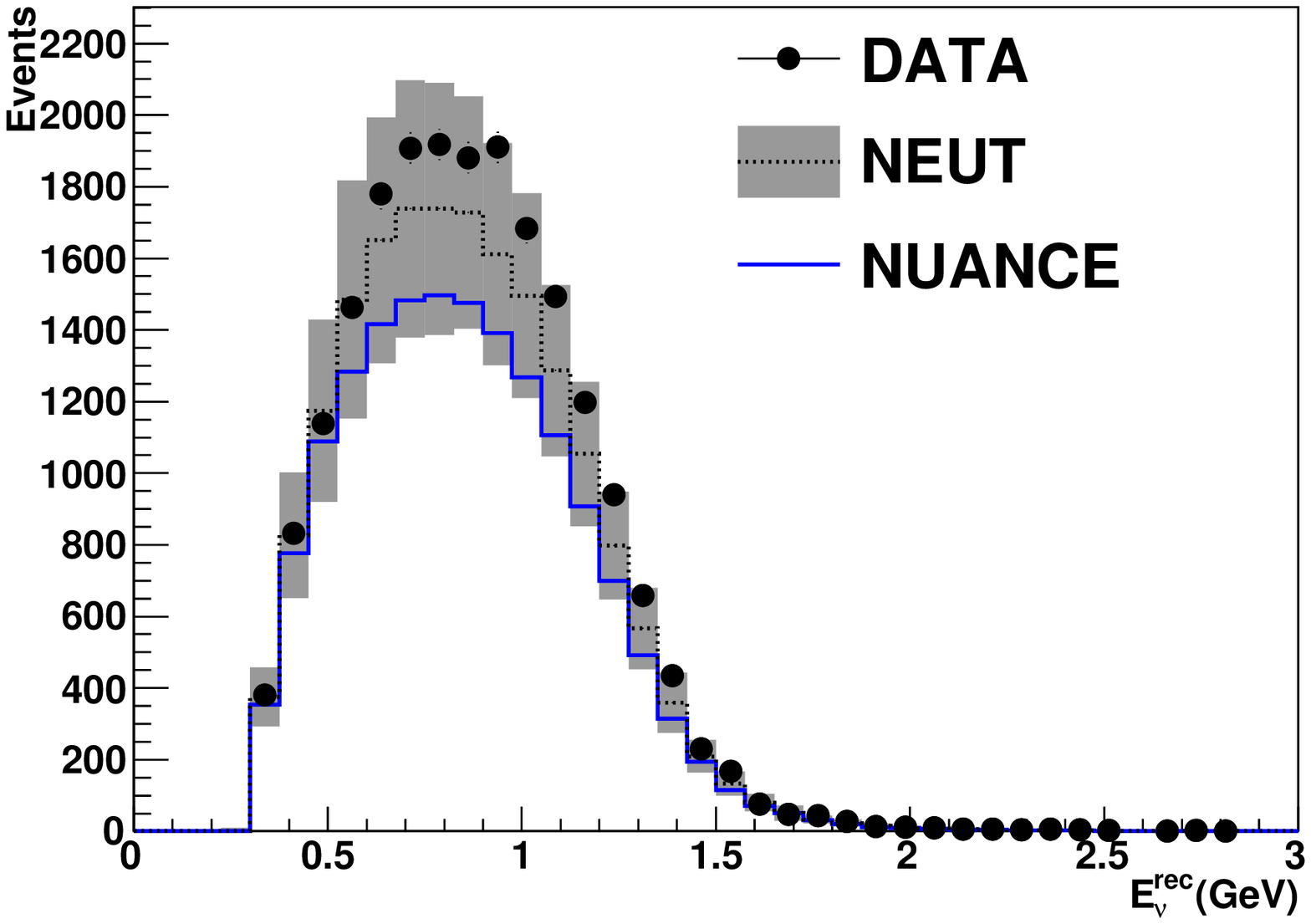}
   \includegraphics[width = \columnwidth]{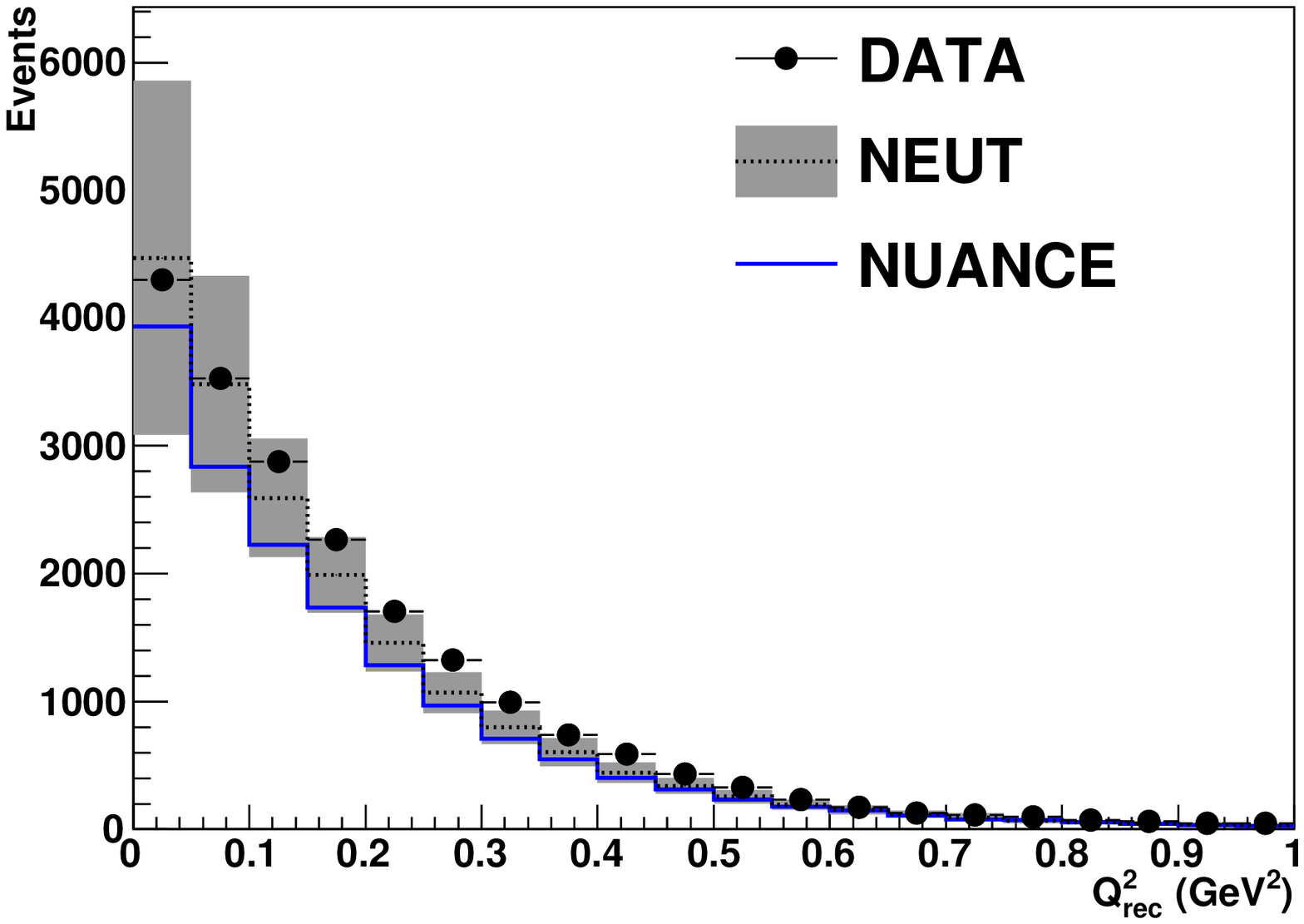}
  \caption{(color online).
    Top: $E_\nu^{rec}$ (left) and $Q^2_{rec}$ (right) of the SciBar-stopped sample.
    Bottom: $E_\nu^{rec}$ (left) and $Q^2_{rec}$ (right) of the MRD-stopped sample.
    The NEUT and NUANCE predictions are absolutely normalized by the number of POT.
    The filled regions show the systematic uncertainties on the MC predictions based
    on NEUT. The systematic uncertainty for the NUANCE prediction is similar to
    that of the NEUT prediction and not shown.
  }
  \label{fig:enu-q2-beforefit}
\end{figure*}

Figure~\ref{fig:enu-q2-beforefit} shows 
the distributions of $E_{\nu}^{rec}$ 
and  $Q^2_{rec}$ for the  SciBar-stopped and MRD-stopped samples.
In these plots, data points are compared with the NEUT and NUANCE based
MC predictions. We find that the data are consistent with the MC predictions
within the systematic uncertainties.

\section{CC Interaction Rate Analysis}
\label{sec:spectrum_fit}

\subsection{Method}
\label{subsec:fit_method}

To calculate the CC inclusive interaction rate and cross section versus energy, 
we re-weight the predictions of  NEUT or NUANCE based simulations
in true energy bins by factors that are found to give 
the best agreement with the kinematic distributions for data versus MC prediction.

\begin{figure*}[htbp]
  \centering
  \includegraphics[width = \columnwidth]{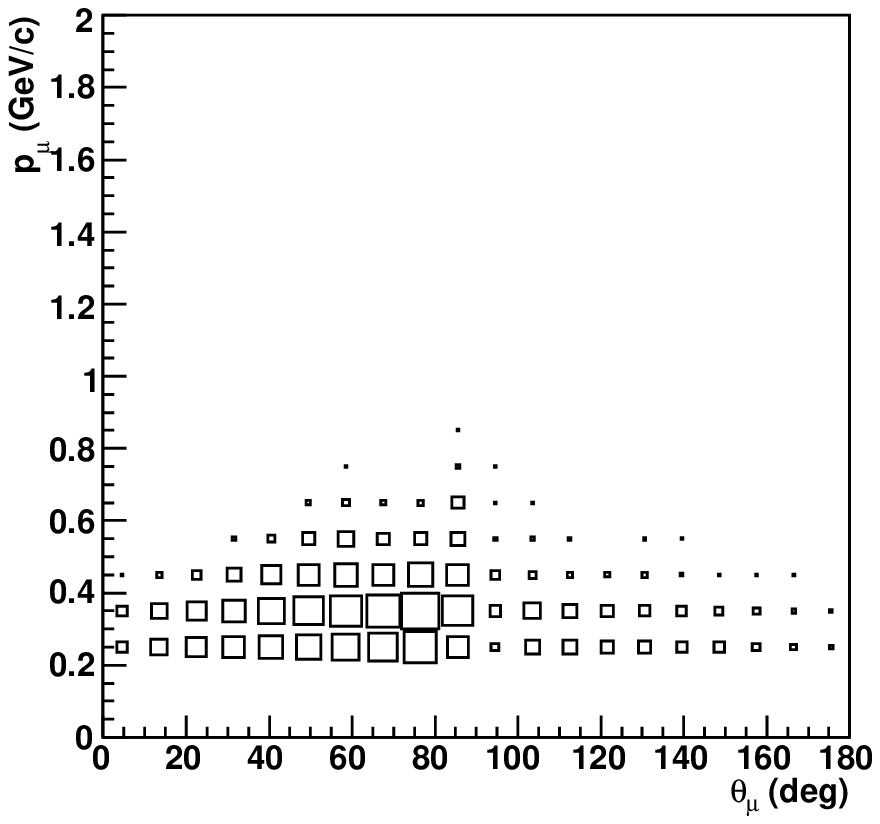}
  \includegraphics[width = \columnwidth]{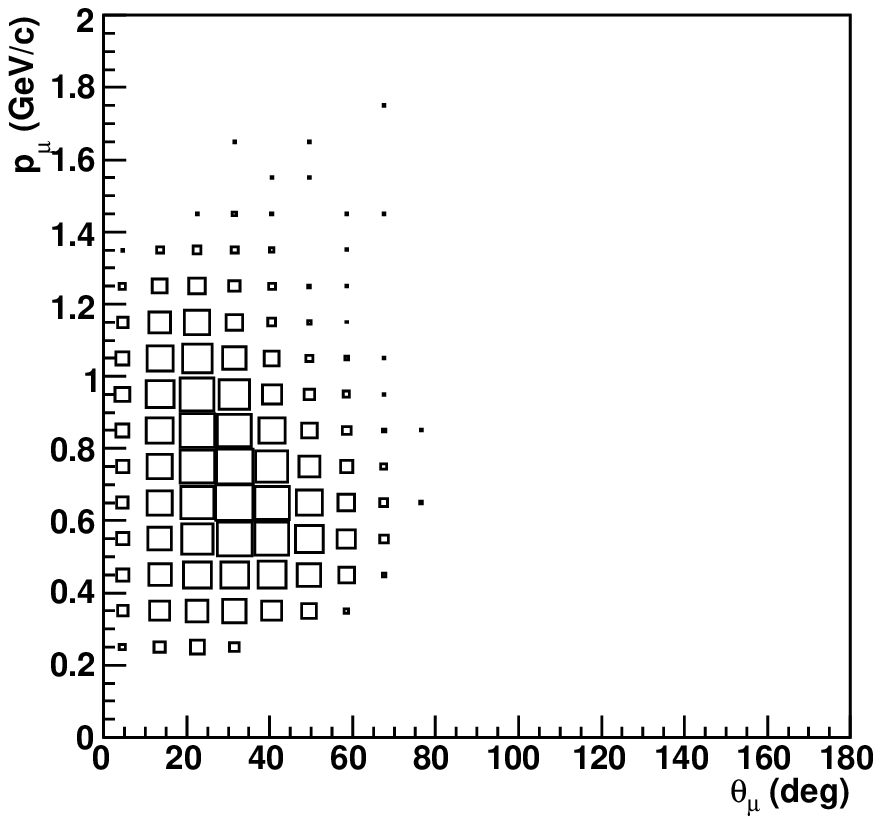}
  \caption{
    Data distributions of $p_{\mu}$ vs. $\theta_{\mu}$ for the SciBar-stopped (left)
    and MRD-stopped (right) samples.
    The size of boxes is proportional to the number of entries.
  }
  \label{fig:mc_pmu_vs_theta}
\end{figure*}

The  $p_\mu$ vs. $\theta_\mu$ ($p_\mu$-$\theta_\mu$) distributions from
the SciBar-stopped and the MRD-stopped samples, and $\theta_\mu$ distribution
from the MRD-penetrated sample are simultaneously used for this measurement.
Figure~\ref{fig:mc_pmu_vs_theta}  shows 
the  $p_\mu$-$\theta_\mu$   distributions of 
the SciBar-stopped and MRD-stopped samples,
while the $\theta_\mu$ distribution for the MRD-penetrated sample 
is shown in Fig.~\ref{fig:pmu-thetamu-pene}.
Events in the same $p_\mu$-$\theta_\mu$ bins but in different sub-samples 
are not summed together, but treated as separate $p_\mu$-$\theta_\mu$
bins in the analysis, and only bins with at least 5 entries are used for the fit.
The total number of $p_\mu$-$\theta_\mu$ bins is 159; 
71 from the SciBar-stopped, 82 from the MRD-stopped and 6 from the MRD-penetrated samples.

\begin{table*}[htbp]
  \centering
  \caption{Energy regions for the CC interaction rate measurement.
  These energy regions are in terms of the true neutrino energy 
  from the MC.}
  \label{tab:fit_parameter_binning}
  \begin{ruledtabular}
  \begin{tabular}{lcccccc}
  Parameter  & $f_0$ & $f_1$ & $f_2$ & $f_3$ & $f_4$ & $f_5$  \\
  \hline
  $E_{\nu}$ range (GeV)&  0.25 - 0.5 &  0.5 - 0.75 & 0.75 - 1.0 & 1.0 - 1.25 & 1.25 - 1.75 & 1.75+ \\
  \end{tabular}
  \end{ruledtabular}
\end{table*}

We  define  6 rate normalization factors ($f_0, \cdots, f_5 $) which represent
the CC interaction rate normalized to the MC prediction
for each true energy
region defined in Table~\ref{tab:fit_parameter_binning}.
The events at $E_\nu < 0.25$~GeV are not used since 
these events are below our detection efficiency as shown in Fig.~\ref{fig:efficiency_enu},
and also the fraction of these low energy interactions
are negligibly small ($< 1\%$) at the BNB flux.
We  calculate
these rate normalization factors 
by comparing the MC predictions to the measured CC interaction rate.
For each energy region, we generate 
the MC templates for the  $p_\mu$-$\theta_\mu$ 
distributions in each event sample; 
$n_{ij}^{pred}$ is the predicted number of events in the j-th 
$p_\mu$-$\theta_\mu$ bin, corresponding to energy bin  $i$.
The expected number of events in each $p_\mu$-$\theta_\mu$ bin, $N_j^{pred}$,
is calculated as
\begin{equation}
\label{eq:mc_template}
N_j^{pred}
= \sum_i^{E_{\nu} bins} f_i n_{ij}^{pred}.
\end{equation}

\begin{figure*}[htbp]
  \centering
  \includegraphics[width = 16cm]{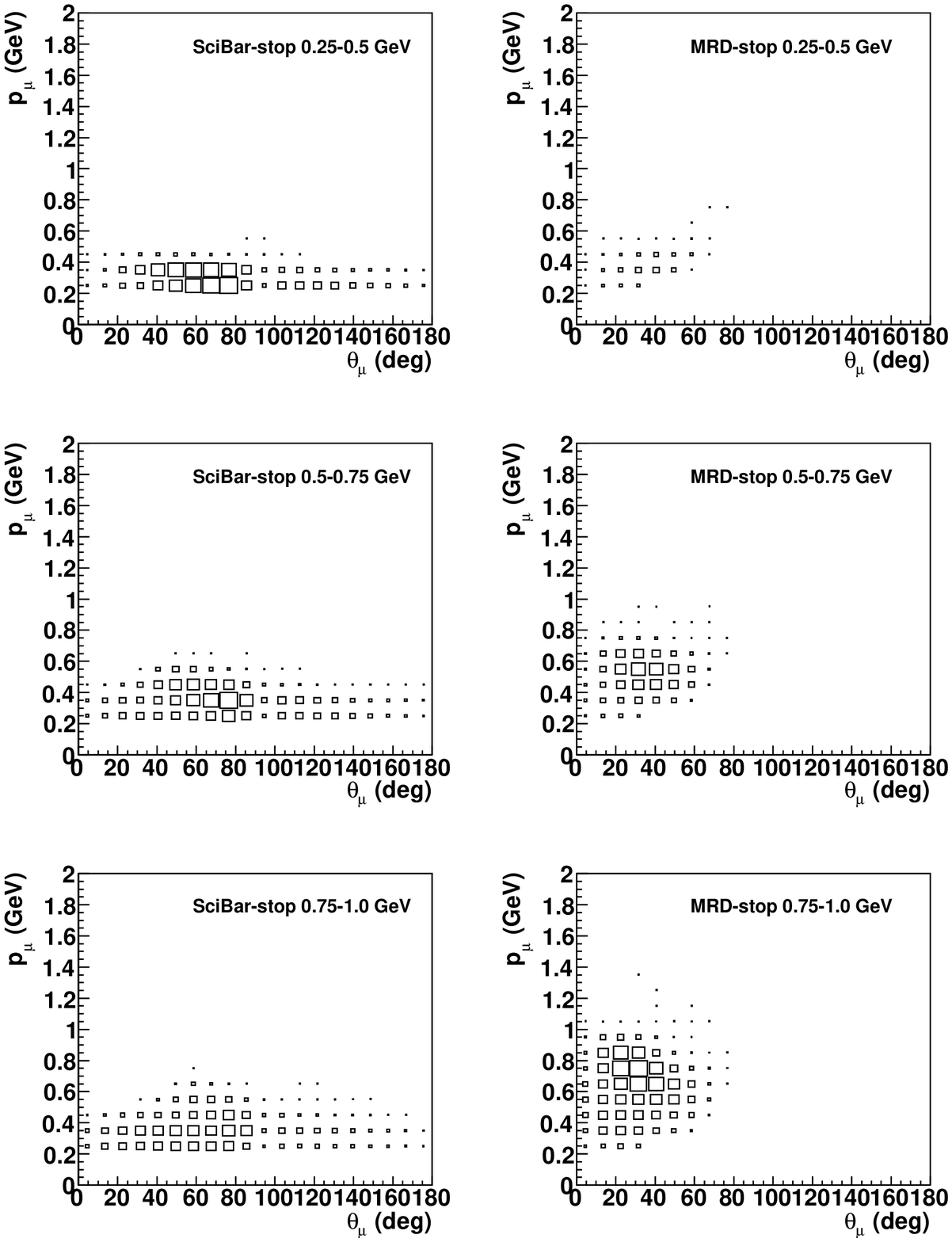}
  \caption{
    The MC templates of $p_{\mu}$ vs. $\theta_{\mu}$ for the SciBar-stopped 
    and MRD-stopped sample for the  three lowest $E_\nu$ regions.
    The normalization factors are common between the
    SciBar-stopped and MRD-stopped samples.
  }
  \label{fig:mc_template_sbmrd_lowe}
\end{figure*}

\begin{figure*}[htbp]
  \centering
  \includegraphics[width = 16cm]{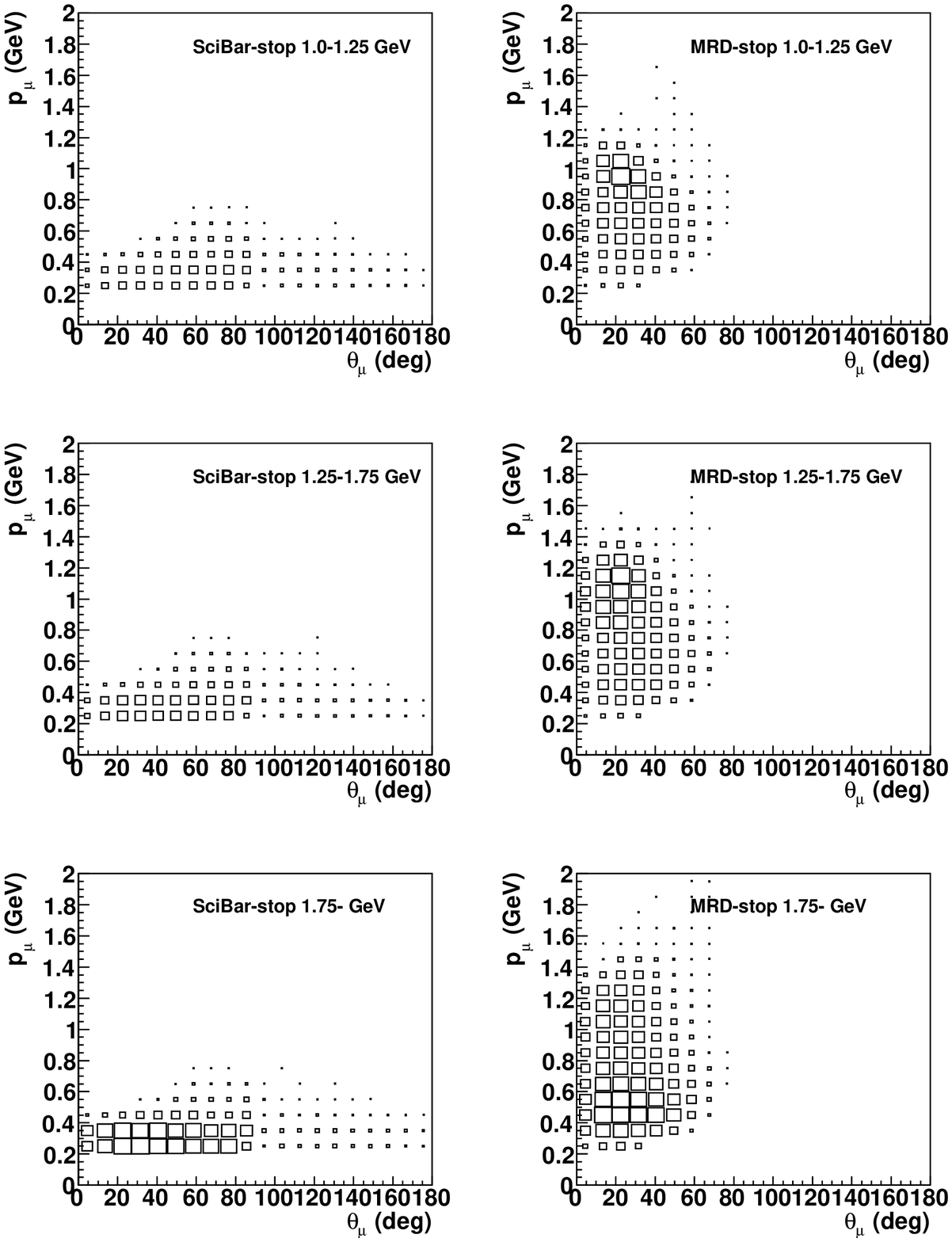}
  \caption{
    The MC templates of $p_{\mu}$ vs. $\theta_{\mu}$ for the SciBar-stopped 
    and MRD-stopped sample for the  three highest $E_\nu$ regions.
    The normalization factors are common between the
    SciBar-stopped and MRD-stopped samples.
  }
  \label{fig:mc_template_sbmrd_highe}
\end{figure*}

\begin{figure}[htbp]
  \centering
    \includegraphics[width = \columnwidth]{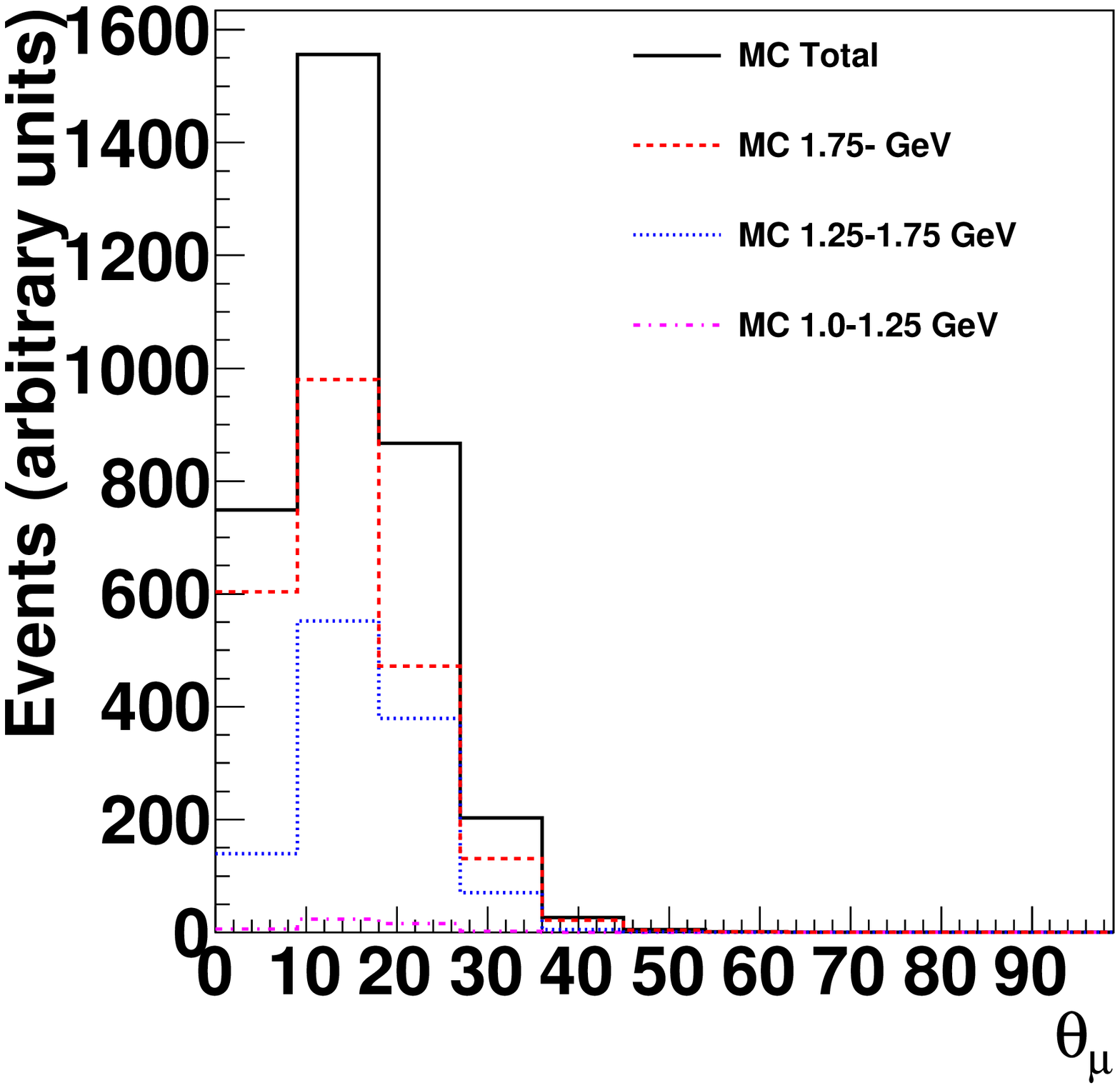}
    \caption{(color online).
      The MC templates of $\theta_{\mu}$ for the MRD-penetrated sample.
      The black line shows the total MC prediction.
      The predictions for $E_{\nu} > 1.75$ GeV,
      $1.25 < E_{\nu} < 1.75$ GeV and
      $1.0 < E_{\nu} < 1.25$ GeV  are also shown.
    }
    \label{fig:mc_template_mrdpene}
\end{figure}

Figures~\ref{fig:mc_template_sbmrd_lowe} and 
\ref{fig:mc_template_sbmrd_highe} are MC templates of the $p_\mu$-$\theta_\mu$ 
distributions for the SciBar-stopped and MRD-stopped samples.
We see that there is a large contribution in the SciBar-stopped sample 
of events with $E_\nu$
below 0.75~GeV. Hence, this sample is essential 
to determine the rate normalization factors in the low energy regions.
The $p_\mu$-$\theta_\mu$ distributions of the
MRD-stopped sample clearly depends on $E_\nu$, up to 1.75~GeV.
However, most of the events in the MRD-stopped sample with $E_\nu>1.75$~GeV have 
small reconstructed $p_\mu$. These are events with
energetic pion or proton tracks  that are mis-reconstructed as muons.
Due to the weak constraint from the MRD-stopped sample on events  with
$E_\nu>1.75$~GeV, 
the MRD-penetrated sample is included in the fit 
since about 2/3 of the events in this sample have $E_\nu>1.75$~GeV
as shown in Fig.~\ref{fig:mc_template_mrdpene}.

We find the rate normalization factors 
($f_0, \cdots, f_5 $) which minimize the
$\chi^2$ value defined as:
\begin{equation}
\label{eq:pearsons-chi2}
\chi^2 = \sum_{j,k}^{Nbins} (N_j^{obs} - N_j^{pred}) 
(V_{sys} + V_{stat})^{-1}_{jk}  (N_k^{obs} - N_k^{pred}).
\end{equation}

Here, $N_{j(k)}^{obs}$ and $N_{j(k)}^{pred}$ are the observed and predicted
numbers of events in the j(k)-th $p_\mu$-$\theta_\mu$ bin, and $N_{j(k)}^{pred}$
is a function of the rate normalization factors as shown in 
Eq.~(\ref{eq:mc_template}).
$V_{sys}$ is the covariance matrix for systematic uncertainties 
in each $p_\mu$-$\theta_\mu$ bin, and
$V_{stat}$ represents the statistical error.
We have a total of 159~bins, so
$V_{sys}$ and $V_{stat}$ are $159 \times 159$ dimensional matrices.
The details of evaluating $V_{sys}$ are described in the following section.

\subsection{Systematic Errors}
\label{subsec:fit_error}

The sources of systematic error are divided into four categories:
neutrino beam (i),
neutrino interaction models (ii),
intra-nuclear interaction model (iii), and
detector response and neutrino interaction models outside of SciBar (iv).
Table~\ref{tab:error-list} shows the list of the systematic uncertainties considered
in this analysis, which were described in the previous sections.

\begin{table*}[htbp]
  \centering
  \caption{List of systematic uncertainties considered.
}
  \label{tab:error-list}
  \begin{ruledtabular}
  \begin{tabular}{clcc}
Category  & Error Source & Variation & Description  \\
  \hline
& $\pi^+/\pi^-$ production from p-Be interaction &  Spline fit to HARP data~\cite{:2007gt} & Sec.~\ref{sec:flux} \\
& $K^+/K^0$ production  from p-Be interaction&  Tables VIII and IX in Ref.~\cite{AguilarArevalo:2008yp} &  Sec.~\ref{sec:flux}\\
(i) 
&  Nucleon and pion interaction in Be/Al  &  Table XIII in Ref.~\cite{AguilarArevalo:2008yp} & Sec.~\ref{sec:flux}\\
 Flux
& Horn current &  $\pm 1$ kA  & Sec.~\ref{sec:flux}\\
& Horn skin effect & Horn skin depth, $\pm$1.4 mm & Sec.~\ref{sec:flux}\\
& Number of POT &  $\pm 2\%$ &  Sec.~\ref{sec:flux}\\
  \hline
  & Fermi surface momentum of carbon nucleus &  $\pm 30$ MeV & Sec.~\ref{sssec:qesiml}\\
  & Binding energy of carbon nucleus & $\pm 9$ MeV &  Sec.~\ref{sssec:qesiml} \\
(ii)   & CC-QE $M_A$ &   $\pm 0.22$ GeV & Sec.~\ref{sssec:qesiml} \\
 Neutrino
& CC-QE $\kappa$ &  $\pm 0.022$ & Sec.~\ref{sssec:qesiml} \\
 interaction
  & CC-1$\pi$ $M_A$ &  $\pm 0.28$ GeV & Sec.~\ref{sssec:spisim} \\
  & CC-1$\pi$ $Q^2$ shape &  Estimated from SciBooNE data & Sec.~\ref{sssec:spisim} \\
  & CC-coherent-$\pi$ $M_A$ & $\pm 0.28$ GeV & Sec.~\ref{sssec:cohpisim}\\
  & CC-multi-$\pi$ $M_A$ & $\pm 0.52$ GeV &  Sec.~\ref{sssec:delsim} \\
  \hline
 & $\Delta$ re-interaction in nucleus & $\pm 100$ \% & Sec.~\ref{sssec:spisim} \\
(iii) & Pion charge exchange in nucleus & $\pm 20$ \% & Sec.~\ref{sssec:fsisim} \\
  Intra-nuclear
 & Pion absorption in nucleus & $\pm 35$ \%& Sec.~\ref{sssec:fsisim}\\
 interaction
 & Proton re-scattering in nucleus & $\pm 10$ \% & Sec.~\ref{sssec:fsisim}\\
 & NC/CC ratio & $\pm 20$ \% & Sec.~\ref{sssec:fsisim} \\
  \hline
  & PMT 1 p.e. resolution &  $\pm 0.20$ & Sec.~\ref{subsec:detector_mc}\\
  & Birk's constant&  $\pm 0.0023$ cm/MeV  & Sec.~\ref{subsec:detector_mc} \\
  & PMT cross-talk&  $\pm 0.004$ & Sec.~\ref{subsec:detector_mc} \\
(iv)   
  & Pion interaction cross section in the detector material&  $\pm 10$ \% & Sec.~\ref{subsec:detector_mc} \\
 Detector
  & dE/dx uncertainty &  $\pm 3$\%(SciBar,MRD), $\pm 10$\%(EC) & Sec.~\ref{subsec:detector_mc} \\
 response
  & Density of SciBar&  $\pm 1$ \% & Sec.~\ref{sec:detector} \\
  & Normalization of interaction rate at the EC/MRD& $\pm 20$ \% & Sec.~\ref{ssec:neuint-overview} \\
  & Normalization of interaction rate at the surrounding materials& $\pm 20$ \% & Sec.~\ref{ssec:neuint-overview} \\
  & Contamination of cosmic-ray backgrounds & Estimated from off-beam data &  Sec.~\ref{subsec:event-classification}\\
  \end{tabular}
  \end{ruledtabular}
\end{table*}

The aim of this fit is to constrain the variations 
due to flux (i) and cross section (ii)
by measuring the rate normalization factors as a function of 
$E_{\nu}$, $f_i$.
Uncertainties from (i) and (ii) are factorized into two parts: 
variations which change the rate normalization 
of CC events in each $E_\nu$ region
 (``normalization'') 
and variations which change the $p_\mu$-$\theta_\mu$ distributions 
but not the normalization of CC events (``$p_\mu$-$\theta_\mu$ shape'').  
The former, normalization, is corrected by the use of $f_i$, 
so $V_{sys}$ corresponds only to $p_\mu$-$\theta_\mu$ shape 
uncertainties for flux and cross section.

The $p_\mu$-$\theta_\mu$ shape uncertainties are estimated by re-normalizing 
the variation for each $E_\nu$ region as follows.
First, we generate a new prediction corresponding to 
a single systematic variation, $n_{ij}'$. 
Here, the prime denotes a systematic variation, 
with $i$ and $j$ representing as before 
the energy bin $i$ and $p_\mu$-$\theta_\mu$ bin $j$.

Then, the predicted event rate, $N_j'$, corresponding to this systematic variation is:
\begin{equation}
\label{eq:shape-variation}
N_j'
= \sum_i^{E_{\nu} bins} f_i n_{ij}'R_i,
\end{equation}
where $R_i$ is the renormalization factor, 
which is the ratio  of the total number of predicted events in each 
$E_\nu$ bin between central value and systematically varied predictions,
defined as:
\begin{equation}
R_i = \frac{\sum_j^{(p_{\mu}, \theta_{\mu}) bins} n_{ij}}{\sum_j^{(p_{\mu}, \theta_{\mu}) bins} n_{ij}'}.
\end{equation}

The predictions for central value $N_j$ and $n_{ij}$  are the same as the 
 $N_j^{pred}$ and $n_{ij}^{pred}$ 
used in Eq.~(\ref{eq:mc_template}).

Unlike the flux and cross section uncertainties, 
intra-nuclear interaction (iii) and detector uncertainties (iv) are 
independent of 
 the CC interaction rate.
For these sources, the uncertainties are simply calculated as:
\begin{equation}
\label{eq:full-variation}
N_j'
= \sum_i^{E_{\nu} bins} f_i n_{ij}'.
\end{equation}

To generate the $V_{sys}$ matrix, the systematically varied rate,
$N_{ij}'$,  is compared to the central value event rate.  
To better estimate some uncertainties, many variation predictions (``draws'') are used, 
each corresponding to a unique set of underlying parameters. 
To estimate the error from the $M_A$ uncertainty, for example, 
we generate 1000 different sets of MC expectations of $n_{ij}'$,
each corresponding to the different value of $M_A$ randomly 
drawn from the estimated uncertainty of $M_A$.
Then, the error matrix representing the uncertainty from $M_A$ is calculated as:
\begin{equation}
V_{jk}^{M_A} = \frac{1}{M}\sum_l^M(N_{jl}' - N_{j})(N_{kl}' - N_{k}),
\end{equation}
where $k$ denotes the index of random draws and $M$ denotes the total number 
of draws.
We estimate the errors from each source (i) - (iv) individually, then add
the matrices together to generate a total covariance matrix, $V_{sys}$.
We assume the same size fractional error for both NEUT and NUANCE 
based predictions, and
the total covariance matrices for them are calculated by scaling the fractional errors
with the predictions.


\section{Results and Discussion}
\label{sec:results}

\subsection{Energy Dependent CC Interaction Rate}
\label{subsec:results-edepend}

\subsubsection{Measurement of CC Interaction Rate}

Tables~\ref{tab:sbspectrum_neut_default}~and~\ref{tab:sbspectrum_nuance_default}
show the 
best fit parameters and their errors from the spectrum fit 
with the six energy dependent rate normalization factors
for NEUT and NUANCE predictions, 
respectively. 
The $\chi^2/DOF$, where DOF is the  number of degree of freedom
for NEUT and NUANCE predictions are respectively
225.1/159 and 560.7/159 before fitting, and they are 
161.2/153 and 173.6/153 after fitting.
We obtain reasonable $\chi^2$ values after the fit.
We also tried to tune $Q^2$ distributions
by fitting cross section parameters such as $M_A$ and $\kappa$, 
however, these give similar $\chi^2$  values.
So we decided to make fits with simple factors  for true $E_\nu$ regions,
as described above,
because they are  less model dependent.

\begin{table}[htbp]
  \centering
  \caption{Spectrum fit result for the  NEUT prediction.
    The first two rows are the best fit values and associated uncertainties.
    The subsequent columns and rows represent the correlation coefficients 
    for each parameters.}
  \label{tab:sbspectrum_neut_default}
  \begin{ruledtabular}
  \begin{tabular}{c|cccccc}
         & $f_0$ & $f_1$ & $f_2$ & $f_3$ & $f_4$ & $f_5$  \\
  \hline
Best fit  & 1.042 & 1.032 & 1.234 & 1.290 & 1.193 & 0.789 \\
Error     & 0.186 & 0.095 & 0.064 & 0.084 & 0.104 & 0.083 \\
  \hline
$f_0$    & 1.0000 & 0.2951 & 0.1653 & -0.2224 & -0.3646 & 0.0438 \\    
$f_1$    & 0.2951 & 1.0000 & 0.0591 & -0.3674 & -0.5311 & -0.2187 \\   
$f_2$    & 0.1653 & 0.0591 & 1.0000 & -0.2317 & -0.1511 & -0.3301 \\   
$f_3$    & -0.2224 & -0.3674 & -0.2317 & 1.0000 & -0.0011 & 0.2534 \\  
$f_4$    & -0.3646 & -0.5311 & -0.1511 & -0.0011 & 1.0000 & -0.0680 \\ 
$f_5$    & 0.0438 & -0.2187 & -0.3301 & 0.2534 & -0.0680 & 1.0000 \\   
  \end{tabular}
  \end{ruledtabular}
\end{table}

\begin{table}[htbp]
  \centering
  \caption{Spectrum fit result for the  NUANCE prediction.
    The first two rows are the best fit values and associated uncertainties.
    The subsequent columns and rows represent the correlation coefficients 
    for each parameters.}
  \label{tab:sbspectrum_nuance_default}
  \begin{ruledtabular}
  \begin{tabular}{c|cccccc}
         & $f_0$ & $f_1$ & $f_2$ & $f_3$ & $f_4$ & $f_5$  \\
  \hline
Best fit   & 1.650 & 1.312 & 1.359 & 1.378 & 1.360 & 0.898 \\
Error      & 0.203 & 0.093 & 0.057 & 0.075 & 0.107 & 0.085 \\
\hline    
$f_0$   & 1.0000 & 0.2260 & 0.0716 & -0.2470 & -0.2916 & 0.0019 \\  
$f_1$   & 0.2260 & 1.0000 & 0.0244 & -0.4319 & -0.5275 & -0.2410 \\ 
$f_2$   & 0.0716 & 0.0244 & 1.0000 & -0.2225 & -0.1150 & -0.3403 \\ 
$f_3$   & -0.2470 & -0.4319 & -0.2225 & 1.0000 & 0.0910 & 0.2003 \\ 
$f_4$   & -0.2916 & -0.5275 & -0.1150 & 0.0910 & 1.0000 & -0.0569 \\
$f_5$   & 0.0019 & -0.2410 & -0.3403 & 0.2003 & -0.0569 & 1.0000 \\ 
  \end{tabular}
  \end{ruledtabular}
\end{table}

\begin{figure*}[htbp]
  \centering
  \includegraphics[width = \columnwidth]{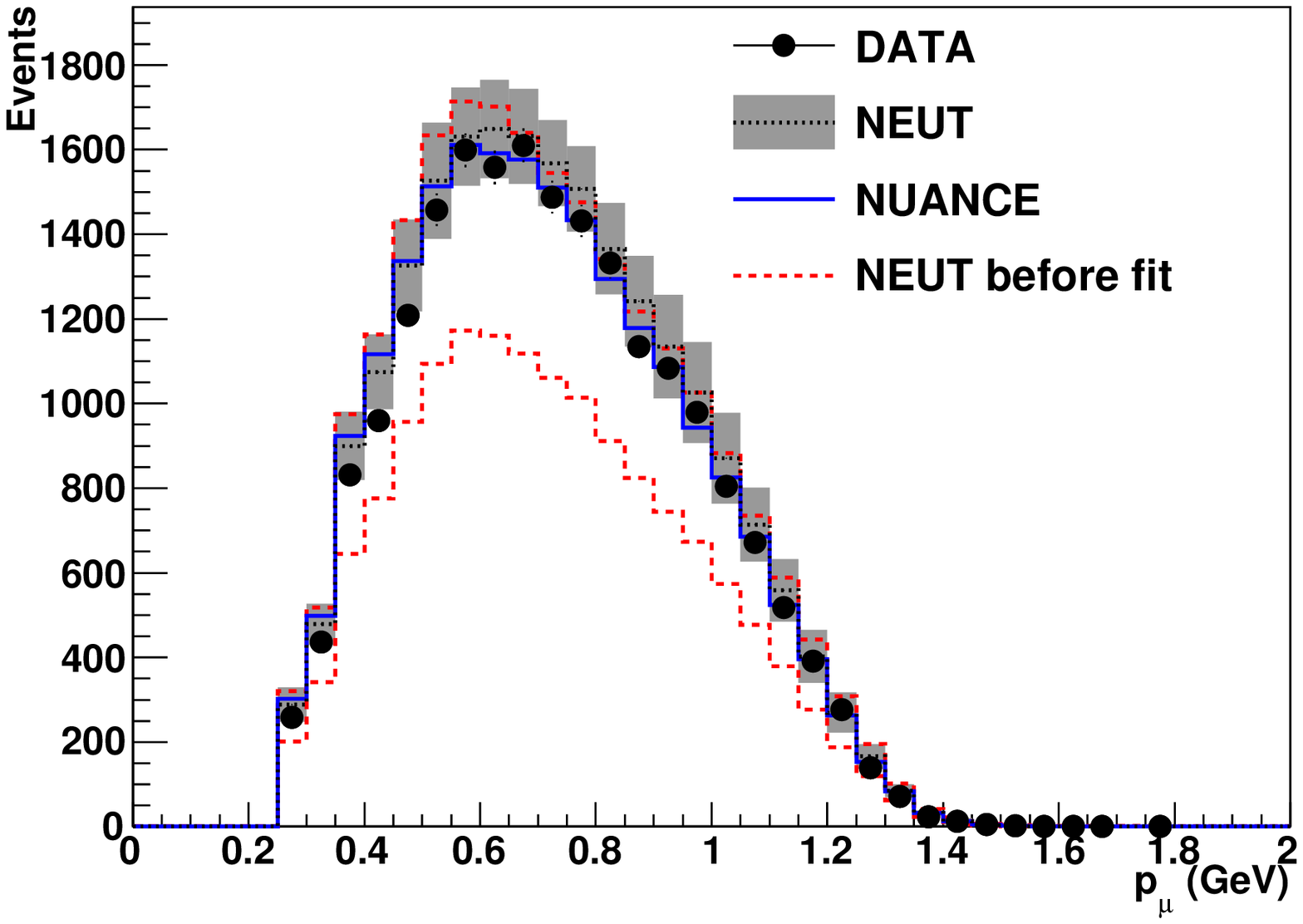}
  \includegraphics[width = \columnwidth]{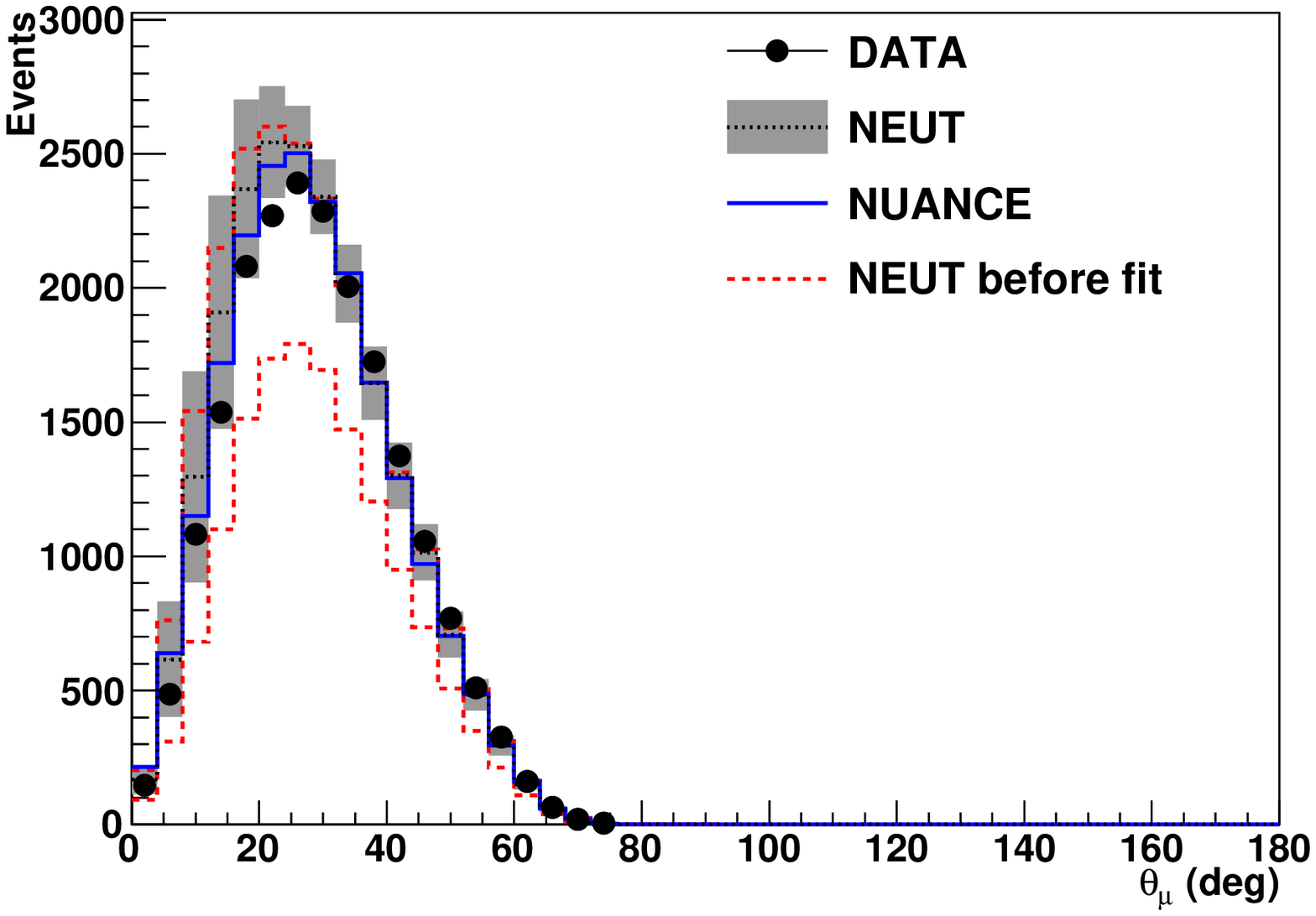}
  \includegraphics[width = \columnwidth]{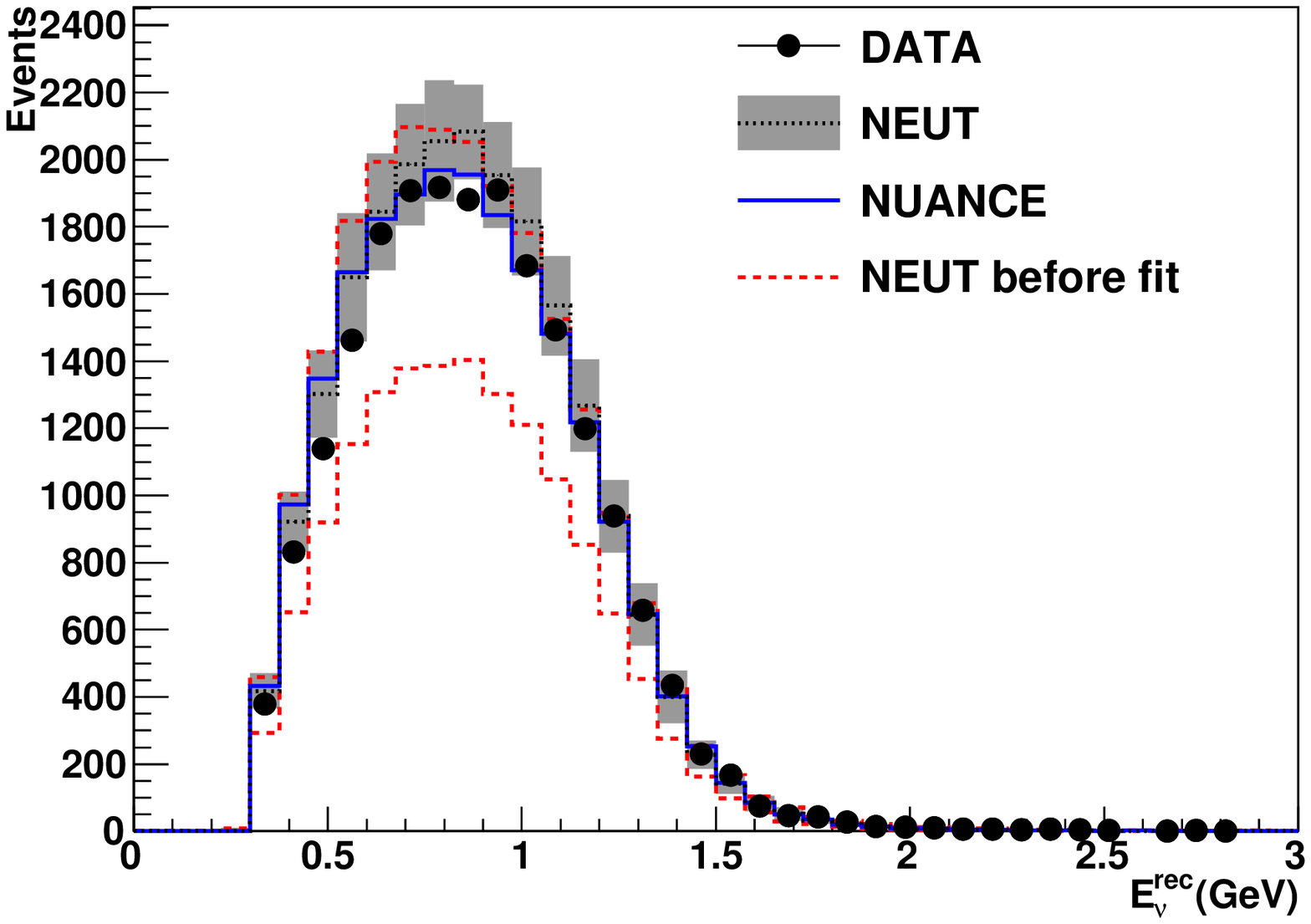}
  \includegraphics[width = \columnwidth]{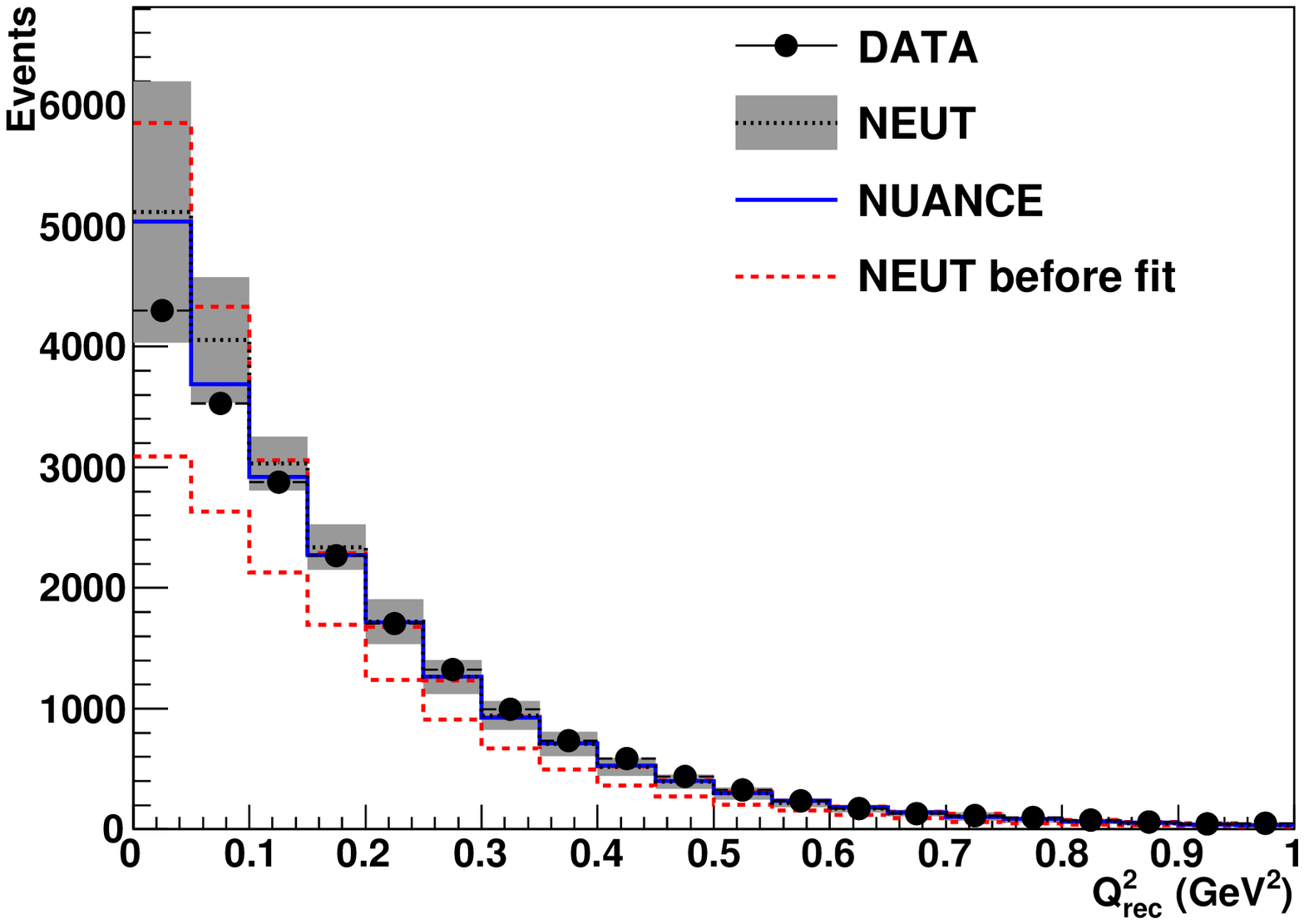}
  \caption{(color online).
    Top: Reconstructed $p_\mu$ (left) and $\theta_\mu$ (right) of the MRD-stopped sample.
    Bottom: $E_\nu^{rec}$ (left) and $Q^2_{rec}$ (right) of the MRD-stopped sample.
    The NEUT and NUANCE predictions after the fit  are shown,
    and the predictions are  absolutely normalized by the number of POT.
    The filled regions show the systematic uncertainties of MC predictions based
    on NEUT. 
    The area within dashed lines shows the NEUT predictions with 
    their systematic uncertainties before the fit.
    The systematic uncertainty for the NUANCE prediction is similar to
    that of the NEUT prediction and not shown.
  }
  \label{fig:enu-q2-afterfit}
\end{figure*}

Figure~\ref{fig:enu-q2-afterfit} shows the distributions of 
$p_{\mu}$,  
$\theta_{\mu}$ 
$E_{\nu}^{rec}$ 
and  $Q^2_{rec}$ of the MRD-stopped sample,
after applying the rate normalization factors obtained in this analysis.
We estimate the constrained systematic error for each distribution 
in the same way as
described in the Sec.~\ref{subsec:fit_error}.
We also propagate the errors of the scale factors ($f_i$) to the distributions.
The errors on $f_i$ obtained from the fit include
the shape error from all the flux and the cross section uncertainties, and 
the absolute error from all the intra-nuclear interaction and detector response uncertainties,
as they are included into the error matrix ($V_{sys}$).
The errors shown in these plots are the quadrature sum of  those constrained
systematic errors and errors of the $f_i$.
We find that both NEUT and NUANCE predictions well reproduce 
the data distributions within the errors of this analysis.
Also, we confirm  that the constraint by this measurement
can reduce the systematic uncertainty
in most regions, 
compared to the original errors.

\begin{figure}[htbp]
  \centering
   \includegraphics[width = \columnwidth]{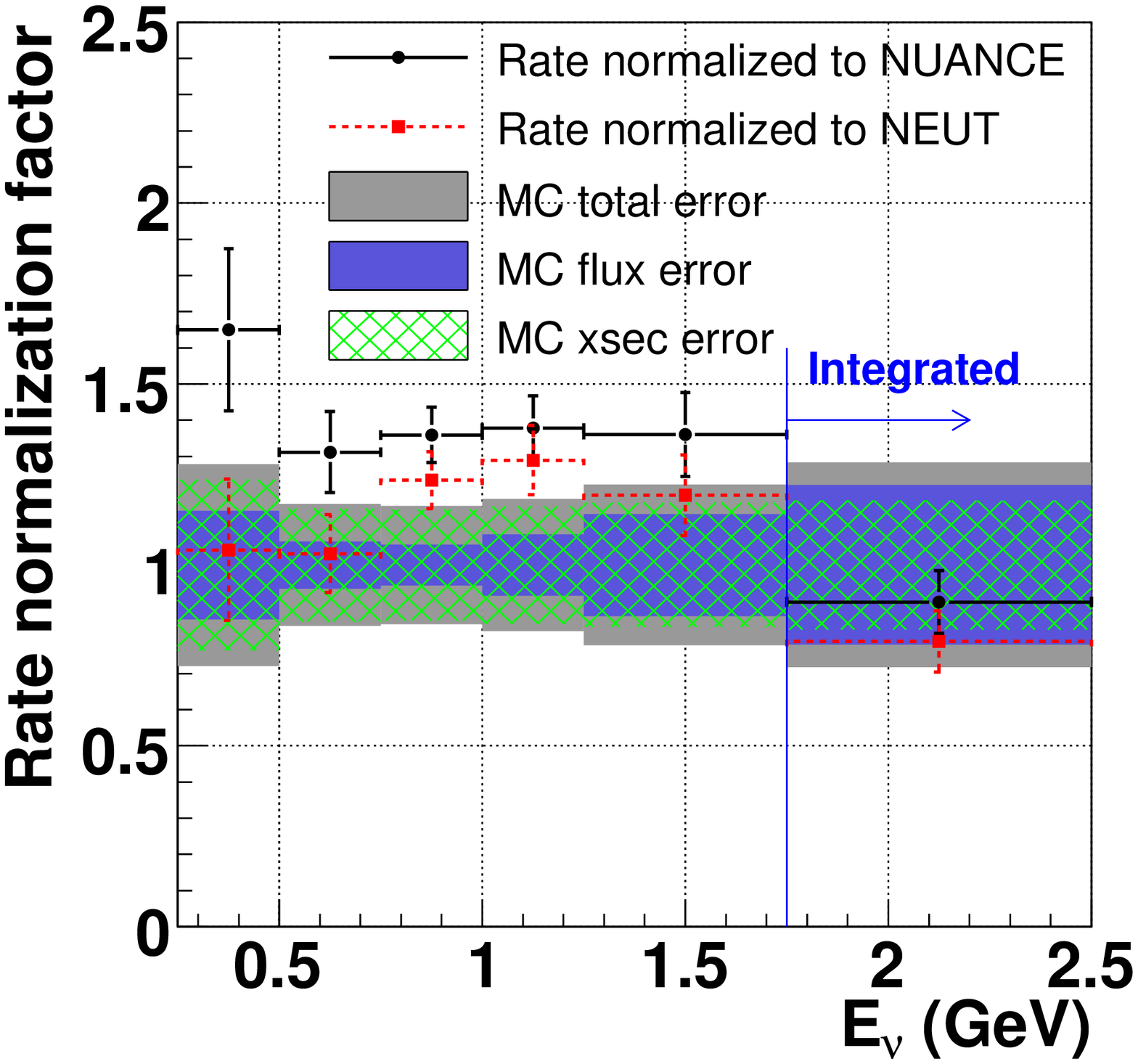}
  \caption{(color online).
    CC interaction rate 
    normalized to the NEUT and NUANCE predictions, 
    obtained by the spectrum fit.
    The error bars show the full systematic uncertainties.
    The original flux and 
    cross section errors are also shown separately.}
  \label{fig:fit-scale-fac}
\end{figure}

The CC interaction rate in i-th true $E_\nu$ region, ${\cal R}_i$,  is 
calculated as:
\begin{equation}
  \label{eq:rate}
  {\cal R}_i = \frac{f_i \cdot {\cal N}^{pred}_{i} \cdot P_i}{\epsilon_i},
\end{equation}
where ${\cal N}^{pred}_i$ is the number of selected events predicted by 
the MC simulation,  $P_i$ is the CC inclusive purity,  
and $\epsilon_i$ is the  CC inclusive efficiency.
We evaluate the errors on $P_i/\epsilon_i$ with all systematic uncertainties
listed  in Table~\ref{tab:error-list}.
Then we take the quadrature sum of the errors of $f_i$ and $P_i/\epsilon_i$ to estimate
the total systematic error.
Figure~\ref{fig:fit-scale-fac} shows 
the obtained neutrino interaction rate 
normalized to the NEUT and NUANCE predictions, with the full systematic error.
The original flux and cross-section uncertainties are also shown in the plot.
The numerical values of the interaction rate 
normalized to the MC predictions and its full errors are 
shown in Table~\ref{tab:cc-inclusicve-edep-rate}.
The uncertainty is about 6\% for the rate at $0.75 < E_\nu < 1.0 $~GeV, where the 
CC interaction rate is maximum, 
and is about 15\% for the lowest energy region.

\begin{table}[htbp]
  \centering
  \caption{$\nu_\mu$ CC inclusive interaction rate normalization factors to
  NEUT and NUANCE predictions. The size of the full systematic errors are 
also shown.}
  \label{tab:cc-inclusicve-edep-rate}
  \begin{ruledtabular}
  \begin{tabular}{ccc}
Energy region  & \multicolumn{2}{c}{$\nu_\mu$ CC rate normalization factor}\\
 (GeV)         & NEUT  & NUANCE \\
  \hline
0.25 - 0.50  &   $1.04 \pm  0.20$  & $1.65 \pm 0.22$ \\
0.50 - 0.75  &   $1.03 \pm  0.11$  & $1.31 \pm 0.11$ \\
0.75 - 1.00  &   $1.23 \pm  0.08$  & $1.36 \pm 0.08$ \\
1.00 - 1.25  &   $1.29 \pm  0.10$  & $1.38 \pm 0.09$ \\
1.25 - 1.75  &   $1.19 \pm  0.11$  & $1.36 \pm 0.12$ \\ 
1.75 -       &   $0.79 \pm  0.08$  & $0.90 \pm 0.09$ \\
  \end{tabular}
  \end{ruledtabular}
\end{table}

\subsubsection{CC Inclusive Cross Section}

The major difference between the rate normalization factors NEUT and NUANCE
is (a) the 
difference in the predicted neutrino interaction cross sections
of the two models.
This effect is compensated by the rate normalization factor 
obtained by this analysis, 
and thus has little effect on the absolute rates.

However, there is also a second-order contribution 
to this difference from  (b) the difference
in the prediction of final state particle production and kinematics.
This affects events with pion or proton tracks 
mis-reconstructed as muons, 
which ultimately affects the purity ($P_i$) and efficiency ($\epsilon_i$) 
in Eq.~(\ref{eq:rate}).

To estimate the effect of (b), 
we extract the absolute CC interaction 
cross section from the obtained rate normalization factors.
Here, the difference between NEUT and NUANCE due to (a) is canceled, 
and the remaining differences are, in principle, due to 
the source (b).
The cross section per nucleon on polystyrene target ($\rm C_8H_8$) at each energy region is calculated as
\begin{equation}
\label{eq:cc-edep-xsec}
\sigma_i
= f_i \cdot <\sigma^{pred}_{CC}>_i 
 = \frac{f_i \cdot {\cal N}^{pred}_{i} \cdot P_i}{\epsilon_i \cdot T \cdot \Phi_i},
\end{equation}
where $i$ is the index of the energy regions used for the spectrum fit
(see Table~\ref{tab:fit_parameter_binning}),
$ <\sigma^{pred}_{CC}>_i $ is the predicted flux averaged CC interaction 
cross section per nucleon, 
${\cal N}^{pred}_i$ is the number of selected events predicted by 
the MC simulation,  $P_i$ is the CC inclusive purity,  
$\epsilon_i$ is the CC inclusive efficiency,
$T_i$ is the number of nucleons in the SciBar fiducial volume, 
and  $\Phi_i$ is the muon neutrino flux per unit area.

\begin{figure}[htbp]
  \centering
   \includegraphics[width = \columnwidth]{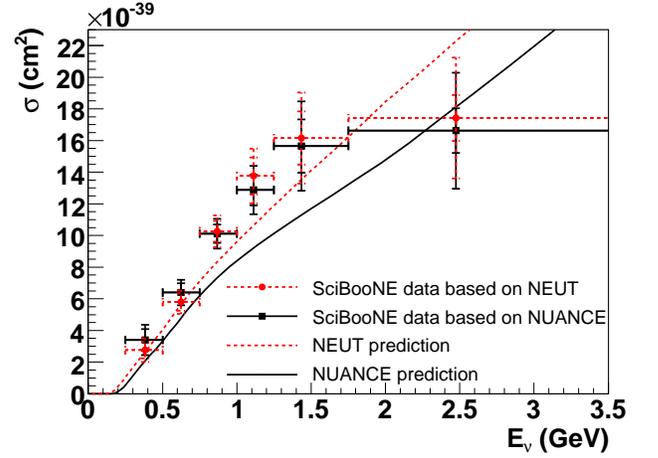}
  \caption{(color online).
    CC inclusive interaction cross section 
    per nucleon on a polystyrene target ($\rm C_8H_8$).
    The smaller error bars show the uncertainties of 
    the rate normalization factors, and 
    the larger error bars represents the total error
    including the flux uncertainties.
  }
  \label{fig:cc-inclusive-xsec}
\end{figure}

Figure~\ref{fig:cc-inclusive-xsec} show the extracted cross sections
plotted with the original predictions from NEUT and NUANCE.
In addition to the errors on $P_i/\epsilon_i$ as estimated for the rate measurements,
we also estimate the errors on $\Phi_i$ from the category (i) in the table.
In the plot, we separately show the errors of $f_i$ and the quadrature 
sum of $f_i$, $P_i$, $\epsilon_i$ and $\Phi_i$ errors.
We confirm that the differences of the extracted CC interaction cross sections
between NEUT and NUANCE are within the errors of $f_i$.
Therefore, the effect of source (b) is small and covered by the 
systematic uncertainty.
The difference of the rate normalization factors is mostly caused by the 
cross section difference itself (source (a)).

The obtained cross section values and their errors are summarized in
Table~\ref{tab:cc-inclusive-edep-xsec}.
The uncertainty on the cross section
is about 10\% at $0.75 < E_\nu < 1.0 $~GeV, where the 
CC interaction rate is maximum, 
and is about 30\% for the lowest energy region.

\begin{table*}[htbp]
  \centering
  \caption{
    Energy dependent CC inclusive cross section per nucleon on a polystyrene target ($\rm C_8H_8$). 
    Results based on NEUT and NUANCE based predictions are separately shown.
  }
  \label{tab:cc-inclusive-edep-xsec}
  \begin{ruledtabular}
  \begin{tabular}{ccccc}
Energy region  & Mean Energy & Total $\nu_\mu$ flux      & \multicolumn{2}{c}{$\nu_\mu$ CC inclusive cross section ($\textrm{cm}^2$/nucleon)}\\
 (GeV)         &   (GeV)     & ($\nu_\mu/\textrm{cm}^2$) & NEUT based & NUANCE based\\
  \hline
0.25 - 0.50  &   0.38  &  $(4.31 \pm  0.81) \times 10^{11}$ &  $(2.76 \pm  0.75) \times 10^{-39}$ &  $(3.40 \pm 0.96) \times 10^{-39} $ \\
0.50 - 0.75  &   0.62  & $(5.09 \pm 0.37) \times 10^{11}$ & $(5.80 \pm  0.75) \times 10^{-39}$ & $(6.39 \pm 0.81) \times 10^{-39} $ \\
0.75 - 1.00  &   0.87  & $(4.18 \pm 0.26) \times 10^{11}$ & $(1.03 \pm  0.10) \times 10^{-38}$ & $(1.01 \pm 0.09) \times 10^{-38} $ \\
1.00 - 1.25  &   1.11  & $(2.63 \pm 0.23) \times 10^{11}$ & $(1.38 \pm  0.17) \times 10^{-38}$ & $(1.29 \pm 0.15) \times 10^{-38} $ \\
1.25 - 1.75  &   1.43  & $(1.90 \pm 0.27) \times 10^{11}$ & $(1.62 \pm  0.29) \times 10^{-38}$ & $(1.56 \pm 0.28) \times 10^{-38} $ \\ 
1.75 -       &   2.47  & $(0.62 \pm 0.12) \times 10^{11}$ & $(1.74 \pm  0.38) \times 10^{-38}$ & $(1.66 \pm 0.37) \times 10^{-38} $ \\
  \end{tabular}
  \end{ruledtabular}
\end{table*}

\subsection{Flux Integrated  CC Interaction Rate}
\label{subsec:results-integrated}
\subsubsection{Measurement of CC Interaction Rate}

We  perform a fit with the method described in Sec.~\ref{sec:spectrum_fit} 
using one rate normalization factor ($f_{tot}$)
which spans the entire energy region, instead of six regions.
This is motivated by the SciBooNE-MiniBooNE joint 
$\nu_\mu$ disappearance analysis, in which we also evaluate the 
sensitivity using one rate normalization factor.

The obtained rate normalization factors are 
$1.12 \pm 0.02 \pm 0.03$, and 
$1.29 \pm 0.02 \pm 0.03$ for the NEUT and NUANCE predictions, respectively.
The first errors show the errors from the fit, and the second  show the errors
from efficiency and purity uncertainties.
Here, higher rate normalization factors 
compared to the simple ratio of the number of events (Table~\ref{tab:nevents}) are obtained.
This is because the ``shape'' systematic uncertainty (Sec.~\ref{subsec:fit_error})
is calculated over the entire energy region, and it typically becomes smallest 
where the CC interaction rate is maximum.
Hence, the obtained rate normalization factors 
tend to be close to the results for $0.75 < E_\nu < 1.0 $~GeV
in the energy dependent analysis
(Sec.~\ref{subsec:results-edepend}).
The $\chi^2/DOF$,
for NEUT and NUANCE predictions are  respectively
208.6/159 and 481.5/159 before fitting, and they are 
173.0/158 and 183.0/158 after fitting.

\subsubsection{CC Inclusive Cross Section}

We extract the total CC inclusive cross section   for the flux at $E_\nu > 0.25$~GeV
using a similar formula to Eq.~(\ref{eq:cc-edep-xsec}):
\begin{equation}
\label{eq:cc-total-xsec}
\sigma 
= f_{tot} \cdot <\sigma^{pred}_{CC}>
= \frac{f_{tot} \cdot {\cal N}^{pred} \cdot P}{\epsilon \cdot T \cdot \Phi}.
\end{equation}
The purity ($P$) and the efficiency ($\epsilon$) are estimated 
for the sum of SciBar-stopped, MRD-stopped and MRD-penetrated samples,
and obtained to be 
$P$ = 90.0(89.1)\% and $\epsilon$ = 34.5(34.5)\%  with the NEUT(NUANCE) generators.
The integrated $\nu_\mu$ flux  at $E_\nu > 0.25$~GeV averaged over the SciBar FV for 9.9E19 POT is 
estimated to be 
$\Phi = (1.87 \pm 0.14) \times 10^{12} ~(\nu_\mu/{\rm cm}^2)$,  
with mean energy of 0.83 GeV.

Using these purity, efficiency and flux predictions, 
we obtain the total CC cross sections 
per nucleon on a polystyrene target 
for a muon neutrino beam with mean energy of 0.83~GeV
to be
$(8.51 \pm 0.72) \times 10^{-39}~\mathrm{cm^2/nucleon}$ and
$(8.45 \pm 0.71) \times 10^{-39}~\mathrm{cm^2/nucleon}$
for NEUT and NUANCE based predictions, respectively.
The error contains all systematic uncertainties from 
the purity and efficiency variation and absolute flux uncertainty.
The error is dominated by the uncertainty on the total flux prediction (7.6\%).
We obtain consistent  CC interaction cross sections
from the two neutrino generator simulations.

\begin{table*}[htbp]
  \centering
  \caption{Extracted number of total CC interaction and absolute cross sections per nucleon on a polystyrene target, based on the 
number of events in each sub-sample. The MC predictions are based on NEUT.
}
  \label{tab:cc_inclusive_xsec}
  \begin{ruledtabular}
  \begin{tabular}{ccccccc}
         & Events      &  Mean energy of & Efficiency   & Purity  & Total CC events      & Flux integrated CC  \\
sample   &  (${\cal N}^{obs}$) & selected sample [GeV] & ($\epsilon$) &  ($P$)&  (${\cal N}^{obs} P/\epsilon$)& cross-section ($\sigma$ [$\rm cm^2$/nucleon])  \\
  \hline
MRD-match & 28409.6 & 1.43 & 27.2\% & 92.9\% & $(9.71 \pm 0.69) \times 10^4$  & $(8.14 \pm 0.85) \times 10^{-39}$  \\
MRD-stop  & 20236.4 & 1.20 & 18.5\% & 91.2\% &  $(10.0 \pm 0.73) \times 10^4$ & $(8.34 \pm 0.88) \times 10^{-39}$  \\
MRD-pene  & 3544.4 & 2.42 &  4.3\% &  97.1\% &  $(7.94 \pm 1.54) \times 10^4$ & $(6.65 \pm 1.29) \times 10^{-39}$  \\
  \end{tabular}
  \end{ruledtabular}
\end{table*}

\subsubsection{CC Inclusive Cross Section for Previous SciBooNE Results}

In previous SciBooNE publications, the total CC interaction cross sections
were used to normalize the results.
We have used the number of MRD-matched and MRD-penetrated samples 
to extract the total CC cross section for CC coherent pion production 
measurements~\cite{Hiraide:2008eu}, 
and used MRD-stopped sample for NC neutral pion production 
measurements~\cite{Kurimoto:2009wq,Kurimoto:2010rc}.
We extract the absolute CC inclusive cross section from the number of 
each sub-sample as:
\begin{equation}
\sigma = \frac{{\cal N}^{obs} \cdot P}{\epsilon \cdot T \cdot \Phi},
\end{equation}
where ${\cal N}^{obs}$ is the observed number of events in each sub-sample, 
$P$ is the purity of $\nu_\mu$ CC interaction in the sample.

Table~\ref{tab:cc_inclusive_xsec} is a summary of the absolute CC cross section
extracted from the number of events in each sub-sample. 
Since we made minor improvements to the MC prediction and the reconstruction algorithms,
the number of events in each sub-sample are slightly different from the previous 
published results.
The difference in the number of events are included in 
 the systematic uncertainties.

For the cross section values extracted from the MRD-matched and the MRD-stopped 
samples, the dominant source of uncertainty is the total flux ($\Phi$) 
uncertainty (7.6\%).
The second largest uncertainty is the efficiency ($\epsilon$) uncertainty
due to the cross section models.
They are 4.0\% and 5.0\% in the MRD-matched and the MRD-stopped samples,
respectively.

For the MRD-penetrated sample, the dominant systematic error from
the efficiency variation is due to the flux model uncertainties.
Approximately 40\% of the MRD-penetrated sample 
is from neutrinos from kaon decay, which have a large uncertainty,
as shown in Fig.~\ref{fig:flux-parents}.
The error from the efficiency variation due to the flux model
 is estimated to be 18.1\%.
The second largest error comes from the
 the total flux ($\Phi$) 
uncertainty (7.6\%).

With these cross section values, one can convert the cross section ratios 
of our previous results to absolute cross sections.


\section{Conclusions}
\label{sec:conclusions}

In conclusion, 
we have isolated three $\nu_\mu$ CC data samples, 
measured $p_\mu$ and $\theta_\mu$, and
extracted the CC interaction rates and cross sections
using flux and neutrino interaction simulations.

We extract the $\nu_\mu$  CC interaction rates
by fitting muon kinematics,
with precision of 6-15\% for the energy dependent
and 3\% for the energy integrated analyses.
We confirm that the distributions after fitting well 
reproduce the observed distributions with both NEUT and NUANCE
based simulations.
This result will be used to constrain the neutrino interaction rate
for a SciBooNE-MiniBooNE joint  $\nu_\mu$ disappearance analysis \cite{Nakajima:2010wc}.

We also evaluate  CC inclusive interaction cross sections, and the 
results are consistent with  both NEUT and NUANCE predictions.
This confirms that the difference in the observed rates 
normalized to the NEUT and NUANCE based predictions is mainly
due to  the cross section difference of the two simulators.
The precisions of the obtained cross sections are 10-30\% for energy dependent
and 8\% for the energy integrated analyses.
This is the first measurement of the CC inclusive cross section on carbon around 1 GeV.
The total CC interaction cross section 
values, as defined in our previous
publications~\cite{Hiraide:2008eu,Kurimoto:2009wq,Kurimoto:2010rc}, 
are also extracted.
These cross section values can be used to convert 
previous measurements of cross section ratios 
of exclusive channels at SciBooNE to the absolute scale.
The results may also be used to tune neutrino interaction 
models in the $\sim$1~GeV region,
which is relevant to various ongoing and future neutrino oscillation experiments.


\section{Acknowledgements}
\label{sec:acknowledgments}

We acknowledge the Physics Department at Chonnam National University,
Dongshin University, and Seoul National University for the loan of
parts used in SciBar and the help in the assembly of SciBar.
We wish to thank the Physics Departments at
the University of Rochester and Kansas State University for the loan
of Hamamatsu PMTs used in the MRD.  We gratefully acknowledge support
from Fermilab as well as various grants, contracts and fellowships
from the MEXT and JSPS (Japan), the INFN (Italy), the Ministry of Science
and Innovation and CSIC (Spain), the STFC (UK), and the DOE and NSF (USA).
This work was supported by MEXT and JSPS with the Grant-in-Aid
for Scientific Research A 19204026, Young Scientists S 20674004,
Young Scientists B 18740145, Scientific Research on Priority Areas
``New Developments of Flavor Physics'', and the global COE program
``The Next Generation of Physics, Spun from Universality and Emergence''.
The project was supported by the Japan/U.S. Cooperation Program in the field
of High Energy Physics and by JSPS and NSF under the Japan-U.S. Cooperative
Science Program.


%

\end{document}